\documentclass[twocolumn]{aastex631}
\shorttitle{UVIT DR1 Catalog}
\shortauthors{Piridi et al.}

\usepackage{amsmath}	% Advanced maths commands
\usepackage{amssymb}	% Extra maths symbols
\usepackage[graphicx]{realboxes}
\usepackage{multirow}
\usepackage{afterpage}
\usepackage{longtable}
\usepackage{supertabular}
\usepackage{comment}
\usepackage{float}
\usepackage{verbatim}
\usepackage{threeparttable}
\usepackage{latexsym}
\usepackage{epsf}
\usepackage[titletoc,title]{appendix}

\begin{document}

\title{A Comprehensive Catalog of UVIT Observations I: Catalog Description and First Release of Source Catalog (UVIT DR1)}

\correspondingauthor{Ananta C. Pradhan}
%
%\email{sonika6195@gmail.com}

\author{Sonika Piridi}
\affiliation{Department of Physics and Astronomy, National Institute of Technology, Rourkela 769008, India}
\email{sonika6195@gmail.com}
 \author{Ranjan Kumar}
 \affiliation{Department of Physics and Astronomy, National Institute of Technology, Rourkela 769008, India}% \\
 \affiliation{Astronomy and Astrophysics Division, Physical Research Laboratory, Navrangpura, Ahmedabad - 380009, Gujarat, India}
 \email{ranjankmr488@gmail.com}
 \author{Divya Pandey}
 \affiliation{Aryabhatta Research Institute of Observational Sciences, Manora Peak, Nainital 263002, India}% \\
 \author{Ananta C. Pradhan}
 \affiliation{Department of Physics and Astronomy, National Institute of Technology, Rourkela 769008, India}
 \email{acp.phy@gmail.com}
%1667 K Street NW, Suite 800 \\
 %Washington, DC 20006, USA}

% \collaboration{20}{(AAS Journals Data Editors)}

% \author{F.X Timmes}
% \affiliation{Arizona State University}
% \affiliation{AAS Journals Associate Editor-in-Chief}

% \author{Amy Hendrickson}
% \altaffiliation{AASTeX v6+ programmer}
% \affiliation{TeXnology Inc.}

% \author{Julie Steffen}
% \affiliation{AAS Director of Publishing}
% \affiliation{American Astronomical Society \\
% 1667 K Street NW, Suite 800 \\
% Washington, DC 20006, USA}

%% Note that the \and command from previous versions of AASTeX is now
%% depreciated in this version as it is no longer necessary. AASTeX 
%% automatically takes care of all commas and "and"s between authors names.

%% AASTeX 6.31 has the new \collaboration and \nocollaboration commands to
%% provide the collaboration status of a group of authors. These commands 
%% can be used either before or after the list of corresponding authors. The
%% argument for \collaboration is the collaboration identifier. Authors are
%% encouraged to surround collaboration identifiers with ()s. The 
%% \nocollaboration command takes no argument and exists to indicate that
%% the nearby authors are not part of surrounding collaborations.

%% Mark off the abstract in the ``abstract'' environment. 
\begin{abstract}

We present the first comprehensive source catalog (UVIT DR1) of ultraviolet (UV) photometry in four far-UV (FUV $\sim$1300$-$1800 \AA) and five near-UV (NUV $\sim$2000$-$3000 \AA) filters of the Ultraviolet Imaging Telescope (UVIT) on board {\em AstroSat}.  UVIT DR1 includes bright UV sources in 291 fields that UVIT detected during its first two years of pointed observation, encompassing an area of 58 square degrees. We used the {\sc ccdlab} pipeline to reduce the L1 data, source-extractor for source detection, and four photometric procedures to determine the magnitudes of the detected sources. We provided the 3$\sigma$ and 5$\sigma$ detection limits for all the filters of UVIT. We describe the details of observation, source extraction methods, and photometry procedures applied to prepare the catalog. In the final UVIT DR1 catalog, we have point sources, extended sources, clumps from nearby galaxies, There are 239,520 unique sources in the combined UVIT DR1, of which 70,488 sources have FUV magnitudes, and 211,410 have NUV magnitudes. We cross-matched and compared non-crowded sources of UVIT with the {\em Galaxy Evolution Explorer (GALEX)} and {\em Gaia} source catalogs. We provide a clean catalog of the unique sources in various UVIT filters that will help further multi-wavelength scientific analysis of the objects.

\end{abstract}
%% Keywords should appear after the \end{abstract} command. 
%% The AAS Journals now uses Unified Astronomy Thesaurus concepts:
%% https://astrothesaurus.org
%% You will be asked to selected these concepts during the submission process
%% but this old "keyword" functionality is maintained in case authors want
%% to include these concepts in their preprints.
% Star clusters (1567); Galaxies (573); Ultraviolet surveys (1742); Nebulae (1095) 

%% From the front matter, we move on to the body of the paper.
%% Sections are demarcated by \section and \subsection, respectively.
%% Observe the use of the LaTeX \label
%% command after the \subsection to give a symbolic KEY to the
%% subsection for cross-referencing in a \ref command.
%% You can use LaTeX's \ref and \label commands to keep track of
%% cross-references to sections, equations, tables, and figures.
%% That way, if you change the order of any elements, LaTeX will
%% automatically renumber them.
%%
%% We recommend that authors also use the natbib \citep
%% and \citet commands to identify citations.  The citations are
%% tied to the reference list via symbolic KEYs. The KEY corresponds
%% to the KEY in the \bibitem in the reference list below. 

\section{Introduction} \label{sec:intro}
The ongoing multi-wavelength space- and ground-based observations continuously explore the puzzling aspects of the Universe. Across radio to gamma-ray wavelength regimes, many contemporary surveys with good depth and resolution have observed vast portions of the sky comprising different classes of objects. Each of these wavelength ranges is crucial to explore multiple aspects of the astrophysical sources. In particular, the ultraviolet (UV) wavelength observations could be effectively used to trace the stellar evolution \citep{2020uvSF}, probe star-forming properties and composition of the interstellar medium (ISM) of galaxies \citep{1994starformation, Calzetti1994, 2011SFgalex, 2011sfrdust}, shed light on the accretion process of active galactic nuclei (AGN) \citep{1999agn, 2001uvxagn, 2023agnastrosat}, etc. However, in the UV wavelength regime, only the {\em Galaxy Evolution Explorer (GALEX)} has conducted a large-scale survey of the sky \citep{Morrissey2007}. Its archival data has been heavily used to study various UV-bright objects and remains the most extensive resource in the UV for planning follow-up observations and missions till now. To facilitate statistical studies of UV sources, \cite{Bianchi2017} and \citet{Bianchi2020} have provided the latest, improved, and expanded versions of the catalogs of all the {\em GALEX} observed sources along with their optical counterparts in {\it Gaia} \citep{Gaia} and {\em Sloan Digital Sky Survey (SDSS)} \citep{sdss}.

With a better resolution and greater sensitivity than {\em GALEX}, the Ultraviolet Imaging Telescope (UVIT) onboard {\em AstroSat}  has been carrying out pointing observations in both far-UV (FUV: 1300 - 1800 \AA) and near-UV (NUV: 2000 - 3000 \AA) for the last eight years and observed simultaneously as long as the NUV detector was operational. Data acquired during the previous eight years show continued good performance of UVIT and is available at the Indian Space Science Data Center (ISSDC).  UVIT is one of the operational telescopes dedicated solely to observing the sky in UV which has the advantage of better resolution and broader field of view (FoV) and continues to expedite significant astronomical results. The UVIT databases will remain one of the comprehensive data sources for studying hot stars and extra-galactic objects such as star-forming galaxies. The other currently observing UV telescopes that have similar features to UVIT are the UltraViolet/Optical Telescope (UVOT) of Swift Observatory, and the XMM-Newton Optical Monitor (XMM-OM).%Therefore, the {\em GALEX} and UVIT databases will remain the most comprehensive and unparalleled data sources for studying hot stars and extra-galactic objects such as star-forming galaxies.

The UVIT observations have contributed to an array of remarkable scientific discoveries since its launch. It has observed several fields pointing towards different Galactic directions of the sky. Around 1600 UVIT fields are publicly available at the {\em AstroSat} archive\footnote{\href{https://astrobrowse.issdc.gov.in/astro_archive/archive/Home.jsp}{https://astrobrowse.issdc.gov.in/astro\_archive/archive/Home.jsp}} for further study. All the data sets can be utilized to extract the source information (e.g., source position, photometric fluxes, size, etc.) and to construct a comprehensive catalog of the UV sources along with object classification (point sources, extended sources, etc.). 

UVIT has established its credibility and uniqueness in observational UV astronomy. While quanta of scientific findings based on UVIT data have already been published, a sizable portion of the UVIT archival observations is yet to be analyzed. Thus, a comprehensive catalog combining sources of different astrophysical origins (galaxies, star clusters, UV-bright hot stars, and other nebular sources) extracted from the archival UVIT observations would widely cater to the scientific community. The photometric information in the catalog could be applicable in performing several statistical studies, e.g., star and galaxy UV luminosity function \citep{2023uvlumfunc} and color-magnitude diagram (CMD) to identify galaxies and stars in various evolutionary phases \citep[][etc]{Divya2021, 2022Pandey,  Kumar2020, ranjan7492}. We analyze the two years of publicly available archival data to provide a source catalog (UVIT DR1) in the FUV and NUV bands of UVIT and report the observed sources along with their UV properties that will further support future investigations. We will discuss the statistical properties of the sources in the catalog and the potential science applications that could be studied using UVIT DR1.

In \autoref{sec:uvit_ins}, we describe the details of the UVIT telescope. In \autoref{sec:data_red}, we describe the data used in UVIT DR1 and the techniques we have used to analyze the data. In \autoref{sec:source_detection}, we explain the procedure to estimate the sky background, the parameters we have used for source detection, and the different kinds of photometry techniques used in the catalog.
In \autoref{sec:validate}, we discuss about the validation of the source catalog and verify UVIT DR1 by cross-matching with {\it GALEX} AIS and {\it Gaia} DR3 catalogs, respectively. In \autoref{sec:cat_detail}, we describe the catalog in detail and the various types of fields observed by UVIT. In \autoref{sec:star-galaxy}, we describe the procedure to classify the probable point and extended sources in UVIT DR1. Finally, we conclude our work in \autoref{sec:summary}.

\noindent The magnitudes of UVIT filters mentioned throughout the paper are in the AB-magnitude system.
	
\section{UVIT Instrument} \label{sec:uvit_ins}
UVIT was launched onboard {\em AstroSat} on September 28, 2015. A detailed instrumentation overview of UVIT can be found in \citet{Kumar2012}, and the ground-based and in-orbit calibrations of UVIT are available in \citet{Postma2011}, \citet{Kumar2012a}, \citet{2017tandon}, and \citet{2020tandon}. UVIT consists of two channels co-aligned with the f/12 Ritchey-Chrétien configuration, each with a 38 cm aperture and a FoV of $28'$. The first channel observes in FUV (1300 - 1800 \AA), whereas the second channel splits the light into two sub-channels using a dichroic beam splitter, and each sub-channel observes in NUV (2000 - 3000 \AA) and visible (3200 - 5500 \AA) passbands, respectively. The FUV, NUV, and visible passbands of the UVIT telescopes consist of five filters each. The details of the imaging filters for each passband are given in \autoref{tab:filter}. CaF2 and BaF2 are the wide-band filters in the FUV channel, and Silica-1 (Silica15) is the wide-band filter in the NUV channel. NUVN2 and Silica have the narrowest bands in the NUV and FUV channels, respectively. Two filters, F148Wa and N242Wa, have never been used for observations.

The in-orbit calibration of UVIT was performed by \citet{2017tandon,2020tandon} using HZ 4, NGC 188, and three overlapping fields in the Small Magellanic Cloud (SMC) with more than 30 months of observations (between December 2015 and October 2018).  It involves the estimation of zero point (ZP) magnitudes, unit conversion factors, point spread function (PSF), flat field corrections, astrometric distortions, etc. They found overall similar performance compared to the ground-based calibrations \citep{Postma2011, Kumar2012a}. The sensitivity of the FUV and NUV filters was 80$-$90\% of the ground-based filter response, except for the NUV N219M filter, which has a reduced sensitivity of $\sim46$\%. The new ZP magnitudes were estimated using HZ 4, which is provided in \autoref{tab:filter} of this paper. The FWHM of PSF of the UVIT filters is less than 1.4$''$ within a diameter of 24$'$. The RMS deviations are found to be $\sim0.3''$ within a diameter of 24$'$ after the astrometric calibrations were carried out using the images in the SMC field.
	
\begin{table}
    \centering
    \caption{Details of the UVIT filters.}
    \label{tab:filter}
    \begin{tabular}{lllllll} 
    \hline
    \hline
        Channel & Filter & Name& ${\rm \lambda_{mean}}$ & ${\rm \Delta\lambda}$ & ZP  \\
		& & & (\AA) & (\AA) & (mag) \\
			\hline
		\multirow{5}{*}{FUV}&	F148W & CaF2-1 & 1481 & 500 & 18.097 \\
		&	F154W & BaF2 & 1541 & 380 & 17.771 \\
            &	F169M & Sapphire & 1608 & 290 & 17.41 \\
		&	F172M & Silica & 1717 & 125 & 16.274 \\
		
		& F148Wa & CaF2-2 & 1485 & 500 & -\\
		\hline
		\multirow{5}{*}{NUV}&	N242W & Silica-1 & 2418 & 785 & 19.763 \\
        &	N219M & NUVB15 & 2196 & 270 & 16.654 \\
		&	N245M & NUVB13 & 2447 & 280 & 18.452 \\
		&	N263M & NUVB4 & 2632 & 275 & 18.146 \\
		
		&	N279N & NUVN2 & 2792 & 90 & 16.416\\
		& N242Wa & Silica-2 & 2418 & 785 & - \\
		\hline
		\multirow{5}{*}{VIS} & V461W & VIS3 &	4614 &	1300 & - \\
		&	V391M &	VIS2 &	3909 &	400	&- 	\\
		&	V347M &	VIS1 &	3466 &	400	&-	\\
		&	V435ND &	ND1	& 4354 &	2200 &- \\		
		&	V420W &	BK7	& 4200	& 2200 & -\\
		\hline
  
\end{tabular}
\end{table}

\section{Observations and Data Reduction} \label{sec:data_red}

\begin{figure*}
    \centering
    \includegraphics[width=0.75\textwidth]{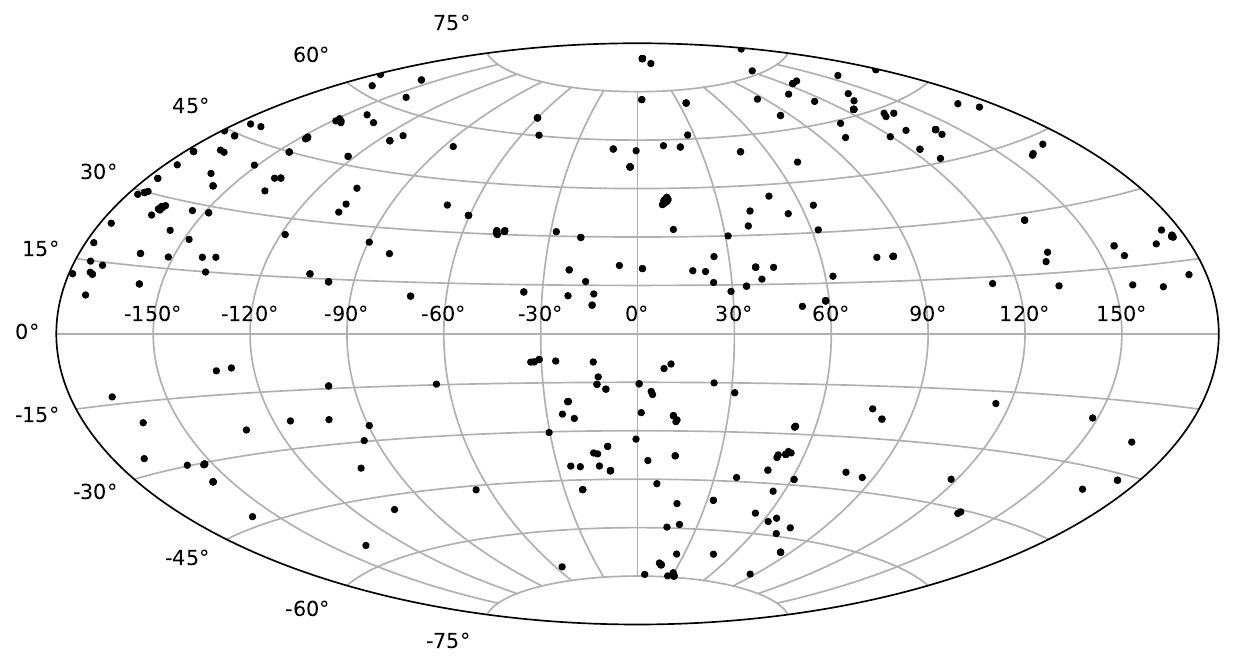}
    \includegraphics[width=0.48\textwidth]{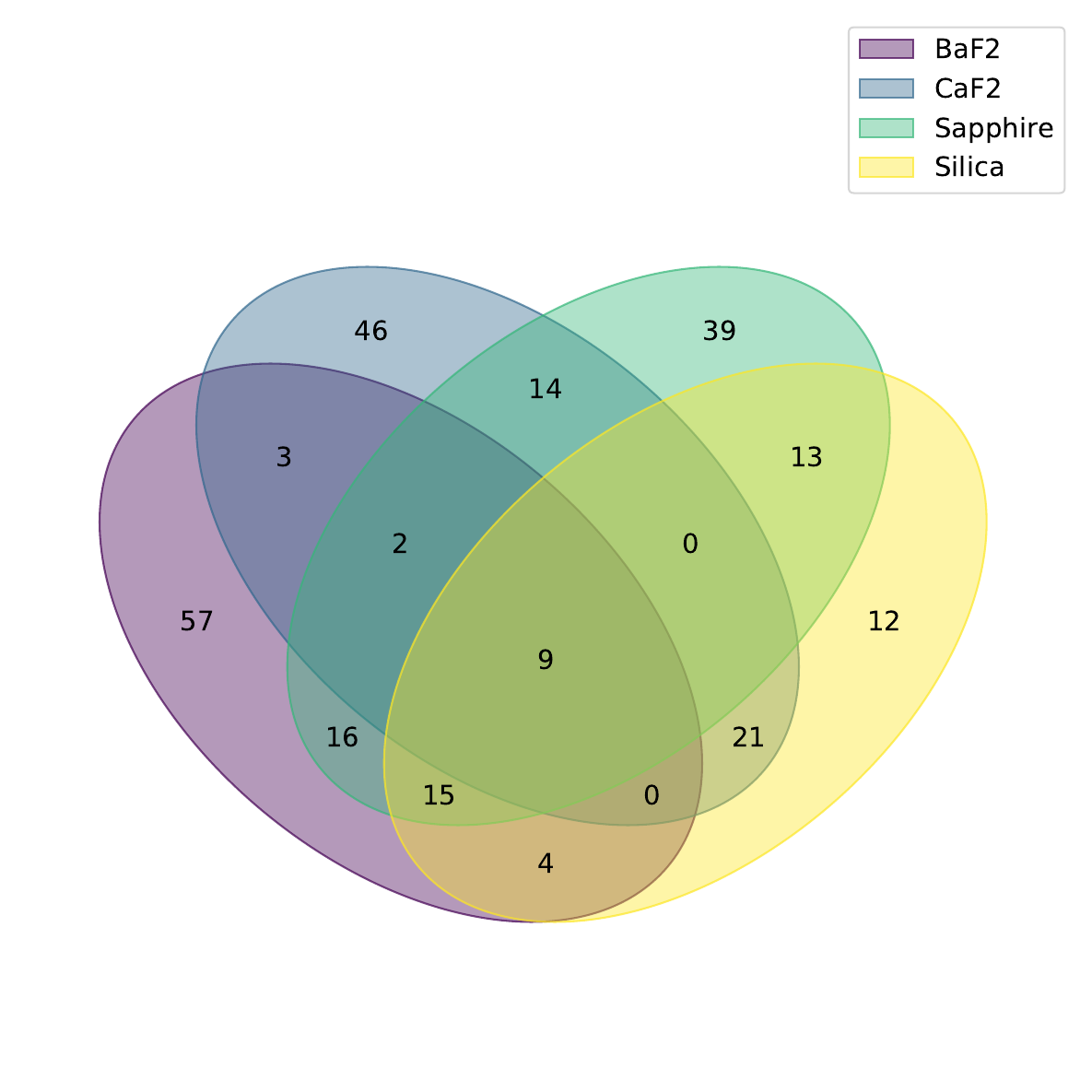}
    \includegraphics[width=0.48\textwidth]{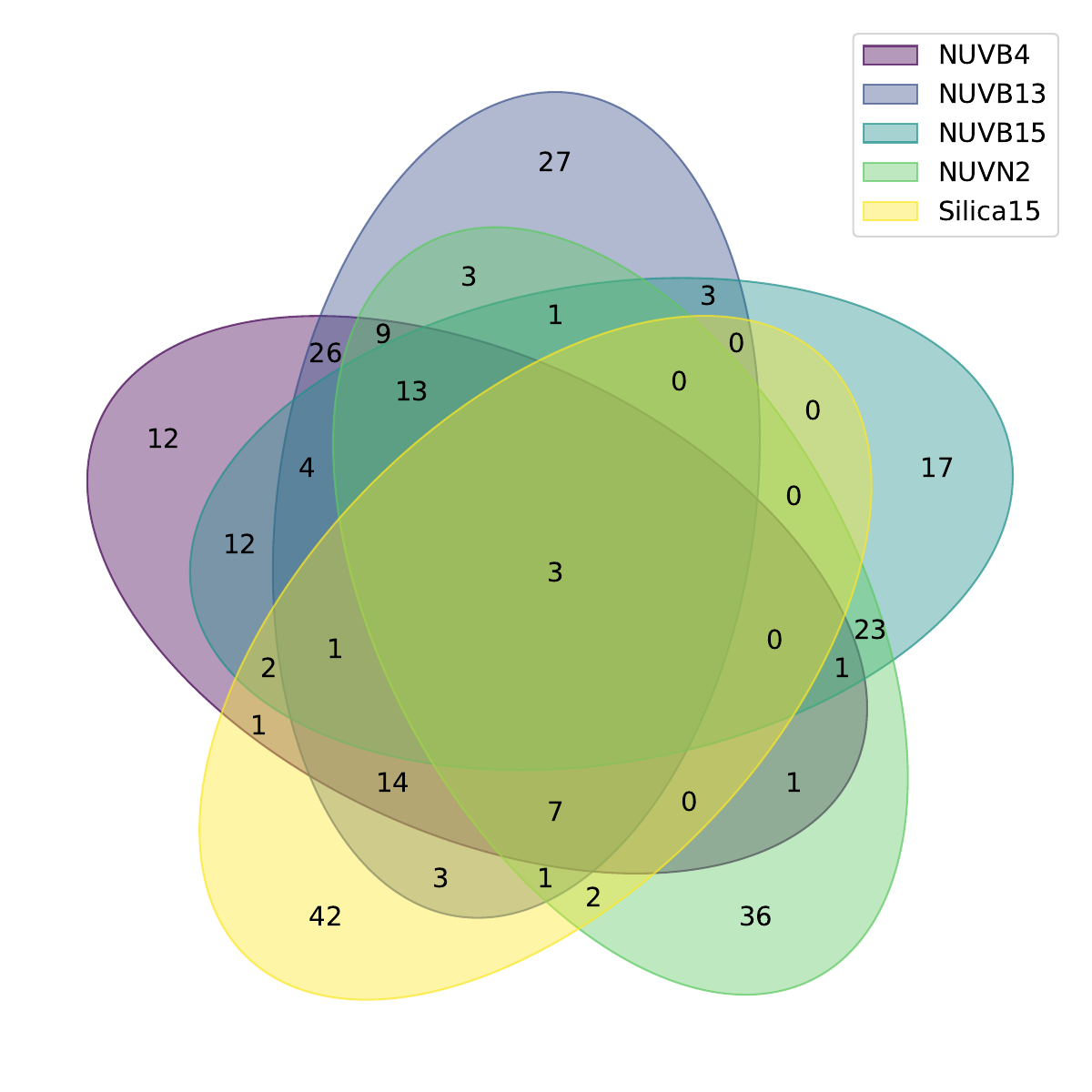}
    \caption{Top panel shows Aitoff Projection of the central celestial coordinates for the 291 fields observed in the year 2016-2017 by UVIT. In bottom panel, Venn diagrams show the distribution of all UVIT observations in different filters. The bottom left panel is for FUV observations, and the bottom right panel is for NUV observations. }
    \label{fig:aitoff}
\end{figure*}

The publicly available L1 data sets of UVIT are archived at the ISRO {\em AstroSat} Archive\footnote{\href{https://astrobrowse.issdc.gov.in/astro\_archive/archive/Home.jsp}{https://astrobrowse.issdc.gov.in/astro\_archive/archive/Home.jsp}}. We downloaded the fields observed by UVIT during the first two years of its observations, starting from ${\rm 1^{st}}$ January 2016 to ${\rm 31^{st}}$ December 2017. We consider the 354 good L1 data files corresponding to as many fields as possible, discarding the ones with no observations in UV filters. Around 44 pointings are observed at multiple epochs within the time period mentioned above. These observations result in 107 L1 files. These L1 files with multiple observations have the same Proposal ID (PID) and Target ID (TID) on {\em AstroSat} archive. To avoid multiple entries of the sources in the catalog, we only include sources from 44 unique L1 files with the longest exposure times. Finally, we work with 291 unique fields, of which 221 have both FUV and NUV observations, 36 have only FUV observations, and 34 have only NUV observations. The exposure times of the observations range from 120 s to 68 ks. All the 291 observed fields are shown in top panel of \autoref{fig:aitoff}. The bottom left and right panels of \autoref{fig:aitoff} show the Venn diagram of the number of observed FoVs in various FUV and NUV filters, respectively. The diagrams also clearly depict a FoV observed by how many filters there are. We use the {\sc ccdlab} pipeline \citep{ccdlab2017, ccdlab2021} for reducing the L1 data. The pipeline produces science-ready images by extracting the L1 files and performing all the telescope calibrations reported in \citet{2017tandon, 2020tandon}.

The zipped L1 files are extracted and separated into FUV, NUV, and VIS directories. The {\em AstroSat} has an equatorial orbit with 90 minutes to complete one orbit. In each orbit, observation is possible for a maximum of 1800 seconds in favorable cases. Hence, it is stored as one orbit (or frame) of observation. For the observations in a filter with an expected exposure time of more than 1800 seconds, the observations are made in several orbits. We have excluded fields with short exposures (less than 120 seconds) as the sources detected in such fields set would have low signal-to-noise (SNR) values and would impact the overall statistics of our catalog. The observed frames in the VIS filters (each with two-sec exposure in the integrated mode) are selected to create a drift series (the difference in positions of each source in the observed frames). The pipeline selects bright sources in VIS images and checks their positions in all the observed frames to create a drift series. When the visual image is very faint, and there are less than 2-3 bright sources in the image, the automatic drift tracking process fails and CCDLAB prompts the user to select the sources manually. We then select the sources manually and check the drift series plot. If the plot is okay (the drift series for all the sources align with each other), we proceed to the next VIS image. If we still cannot find a good source, we create the image using the NUV/FUV drift series. If even that fails, we discard the image. The drift series obtained from the VIS filters is applied to the FUV and NUV filters to track and correct the drift in their respective frames. The drift-corrected frames of the FUV/NUV filters were then added to create 2D images for individual orbits. After orbit-wise drift correction, we select a few bright sources (at least 3$-$4 in FUV and 4$-$5 in NUV) in each orbit to align and register all the orbits in each UVIT filter. All the orbit-wise images of the FUV and NUV filters are aligned with respect to the selected source positions. After the alignment of all the images, we merge all the orbit-wise images to generate a combined master image in the respective filters of FUV and NUV channels with an effective exposure as a sum of the exposure times of all the observed orbits. We adopt an additional `optimization' technique available in {\sc ccdlab} to remove the artificially created streak-like features. These slightly extended streak-like features in the master images often arise due to inaccurate drift tracking in the VIS channel \citep{ccdlab2021}. We obtain an exposure map and master image for each filter as an output.

{\sc ccdlab} uses the {\em Gaia} DR2 catalog \citep{gaiadr2} to apply the world coordinate system (WCS) solution to the master images. The central celestial coordinates of the field, right ascension (RA), and declination (DEC) values from the header of the master images are used to download a reference {\it Gaia} DR2 source catalog within a region of $\sim$17$'$ radius from the given central RA and DEC values. The astrometric solutions determined with {\it Gaia} DR2 are then matched to a set of (at least four) sources detected in the image. We give the `stopping criteria' to be either 8 and decrease it up to 3 points or 75\% of the total number of sources in the catalog. The astrometric solution is assumed to be good if either of these conditions are satisfied. The uncertainties in astrometry lie between 0.1$''$ and 0.5$''$\footnote{If the WCS solution is not found within this limit, we use the CCMAP package in IRAF to find the WCS solution manually.}. After satisfying the astrometric correction, we finalize the observed science-ready image into a zipped file. The final science-ready images include different astrophysical sources, i.e., nearby galaxies, planetary nebulae, and open and globular clusters.

\section{Source Detection and Photometry} \label{sec:source_detection}

\subsection{Sky background and magnitude detection limits of the UVIT images} \label{sec:sky_bkg}

\begin{figure*}
    \centering
    \includegraphics[width=1.0\columnwidth]{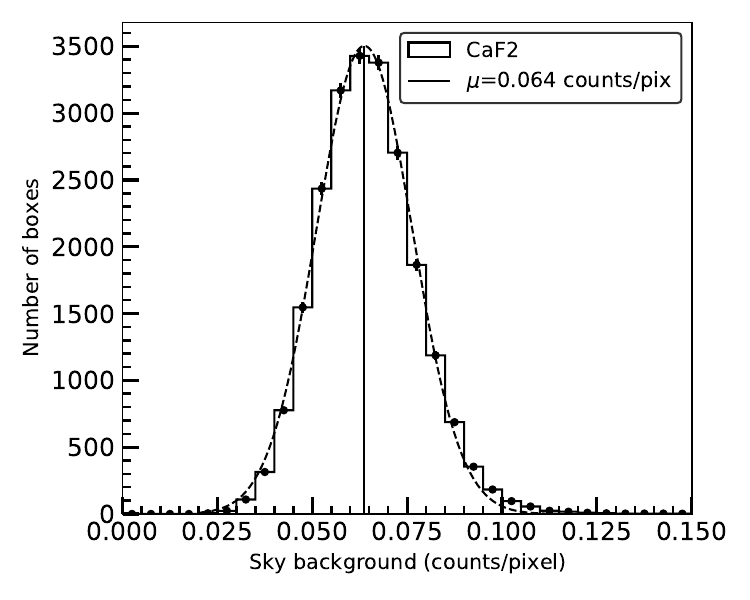}
    \includegraphics[width=1.1\columnwidth]{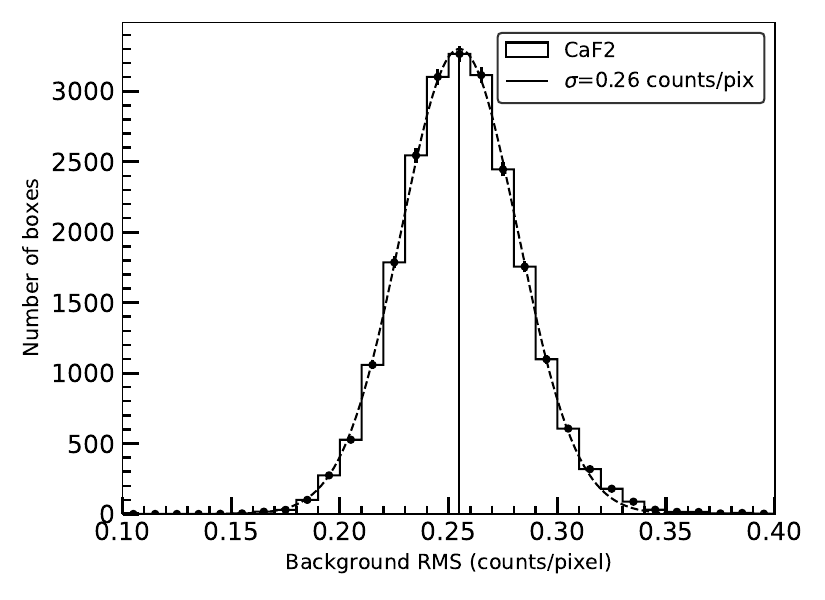}  
    \caption{We fit the Gaussian functions on the histograms of the mean $\mu$ (left panel) and mean $\sigma$ (right panel) calculated for each 21 $\times$ 21 pixels box in an image. The vertical lines display mean $\mu$ and mean $\sigma$ values for a FUV observation determined from the Gaussian fits. 
    } 
    \label{fig:sky-bkg}
\end{figure*}

Science-ready images produced by the standard data reduction have not had the sky background subtracted. Hence, we estimate the background for each observation computed using the random box method. We use the segmentation map generated from source-extractor \citep[SE,][]{sextractor} to mask sources above a threshold of 2$\sigma$ from the background. We consider boxes of size $21\times 21$ pixels on the masked image with the criteria that at least 1000 boxes must be present throughout the whole image for sky estimation, avoiding the masked pixels. If the total number of boxes in an image is less than 1000, we reduce the box size in descending order from $15\times 15$ pixels, $11\times11$ pixels, $7\times 7$ pixels to $5\times 5$ pixels to ensure the total number of boxes above 1000. This is important because it enables the assessment of statistical uncertainty in the background measurements. However, we were able to estimate the background with 21 $\times$ 21 pixels box size for all the fields.

We estimated the background mean ($\mu$) and the background RMS error ($\sigma$) values for each box. The histogram plot, \autoref{fig:sky-bkg}, shows the distribution of the observed $\mu$ and $\sigma$ obtained in each box for an image in the CaF2 (exposure: 2900 s) filter for a UVIT field having PID:A03\_036 and TID:T01. We fitted a Gaussian function over the plotted histograms for estimating $\mu_{\rm mean}$ and $\sigma_{\rm mean}$ for the CaF2 filter as 0.064 counts/pixel and 0.26 counts/pixel, respectively. The procedure was repeated for all the UVIT observed fields. The mean sky background is subtracted from each image before photometry.

We calculate the UVIT detection limit at 3$\sigma$ and 5$\sigma$ within a circular aperture of radius 1.0$''$ using \autoref{eq:detect_limit} as mentioned in \citep[][etc]{2023mondal, 2020kanak}. 

\begin{equation} \label{eq:detect_limit}
 {\rm m_{3\sigma ( 5\sigma)} = -2.5 \times log_{10}[ 3\sigma (5\sigma) \times \sqrt{N_{pix}}] + ZP}\\
\end{equation}

where ${\rm N_{pix}}$ is the number of pixels in the circular aperture. 

A linear variation between the limiting magnitudes ($m_{3\sigma}$ and $m_{5\sigma}$) and the logarithm of the exposure times is noticed for all the UVIT filters. We fit a logarithmic function, 
\begin{equation} \label{eq:log_fitting}
m = A + B \times \log_{10}( T_{exp})
\end{equation}

where $m$ is the magnitude detection limit, T$_{exp}$ is exposure time, and A and B are fitting coefficients. %The fitted lines for the $3\sigma$ and $5\sigma$ detection limits are shown in black and red solid lines, respectively, in \autoref{fig:exp-mag-lim}.
We provide the estimated values of the A and B coefficients in \autoref{tab:mag-lim-fit}, which can be used to determine the $3\sigma$ and $5\sigma$ limiting magnitudes for any UVIT filter for a specific exposure time. We also provide the typical 5$\sigma$ magnitude detection limit ($m_{5\sigma}$) for an observation of 100 s for all the filters of UVIT in \autoref{tab:mag-lim-fit}, and these values tend towards the fainter magnitudes for higher exposure times. The estimated $m_{5\sigma}$ for an exposure time of $\sim$100 s in the wide-band filter Silica15 (BaF2) of UVIT is 26.7 (25.6) mag. Note that for comparison the typical 5$\sigma$ detection limit of the {\em GALEX} AIS survey for NUV (FUV) filter for an exposure time of 100 s is $\sim$21 (20) mag \citep{Bianchi2017}.

\begin{table*}
    \centering
    \caption{Fitting coefficients of the \autoref{eq:log_fitting} for $3\sigma$ and $5\sigma$ magnitude detection limits of UVIT filters {\bf measured within a circular aperture radius of 1.0$''$}. $N_{F}$ is the number of fields observed in each filter. ${ \rm FWHM_{PSF}}$ is the mean of the PSF FWHM of the PSF of all the fields in each filter. $m_{5\sigma}$ is the typical 5$\sigma$ magnitude detection limit for an observation of 100 s. }
    \label{tab:mag-lim-fit}
    
    \begin{tabular}{lccccccr}
    \hline
    \hline
        Filter &	$A_{3\sigma}$	&	$B_{3\sigma}$ &	$A_{5\sigma}$ &	$B_{5\sigma}$ & m$_{5\sigma}$ & ${ \rm FWHM_{PSF}}$ & $N_{F}$\\
               &	mag 	&	mag &	mag & mag &	mag & arcsec & \\
    \hline
        CaF2 & 24.1 $\pm$ 0.2 &	1.25 $\pm$	0.05 &	23.5 $\pm$	0.2 &	1.25 $\pm$	0.06 & 26.0 & 1.40 $\pm$ 0.38 & 95\\
        BaF2 & 23.4 $\pm$ 0.2 &	1.36 $\pm$	0.05 &	22.8 $\pm$	0.2 &	1.36 $\pm$	0.05 & 25.6&  1.41	$\pm$ 0.41 & 106\\
        Sapphire & 23.1 $\pm$ 0.2 &	1.35 $\pm$	0.07 &	22.5 $\pm$	0.2 &	1.35 $\pm$	0.07 & 25.2 & 1.44 $\pm$ 0.45 & 109\\
        Silica & 22.7 $\pm$	0.2 & 1.2 $\pm$ 0.05 &	22.2 $\pm$	0.2 &	1.2 $\pm$	0.05 & 24.5 & 1.40	$\pm$ 0.35 & 75\\
        
        Silica15 & 24.7 $\pm$ 0.1 &	1.27 $\pm$	0.03 &	24.1 $\pm$	0.1 &	1.27 $\pm$	0.03 & 26.7 &1.04 $\pm$ 0.18 & 76\\
        NUVB15 & 22.9 $\pm$ 0.1 & 	1.21 $\pm$	0.02 &	22.3 $\pm$ 	0.1 &	1.21 $\pm$	0.02 & 24.7 &1.31	$\pm$ 0.28 & 80\\
        NUVB13 & 24.1 $\pm$ 0.1 &	1.24 $\pm$	0.03 &	23.5 $\pm$ 0.1 &	1.24 $\pm$	0.03 & 26.0 &1.14	$\pm$ 0.35 & 115\\
        NUVB4 &	23.6 $\pm$	0.1 & 1.24 $\pm$	0.03 &	23.1 $\pm$	0.1 &	1.23 $\pm$	0.03 &  25.5 &1.08	$\pm$ 0.34 & 106\\
        NUVN2 &	24.7 $\pm$ 0.1 & 1.28 $\pm$	0.03 &	24.1 $\pm$	0.1 &	1.28 $\pm$	0.03 & 24.2 & 1.04	$\pm$ 0.24 & 100\\
    \hline
    \end{tabular}
\end{table*}

\subsection{Point Spread Function of the UVIT images} \label{sec:psf}

We use Point Source Function Extractor\footnote{ PSFEx is an extension of the SE specifically designed to estimate the PSF of the point-like sources in the image. \href{https://psfex.readthedocs.io/en/latest/index.html}{https://psfex.readthedocs.io/en/latest/index.html}} \citep[PSFEx;][]{Bertin2011} to estimate the FWHM of PSF in the FUV and NUV filter images of the observed fields. With a detection threshold of 10$\sigma$, we run SE on the science-ready images to obtain a list of bright sources. A few parameters acquired as an output from SE, such as ellipticity and FWHM, are used to identify the probable point sources in each field, which are further fed to {\mbox PSFEx}. SE computes the FWHM of each source, assuming the source follows a Gaussian profile, giving a rudimentary idea about their PSF width. We include in the catalog the sources with SE FWHM between 2 (0.8$''$) and 10 pixels (4.17$''$) and ellipticity $<$ 0.18 to PSFEx.

PSFEx models the PSF as a linear combination of basis vectors \citep{2011psfex}. The output file from PSFEx contains a table that lists the number of sources used to model the PSF, mean PSF FWHM, ellipticity, and $\chi^2$. The resultant PSF FWHM for FUV and NUV filters were found to be $1.0'' - 2.0''$ and $0.8'' - 1.5''$, respectively, except for a few outliers over the ranges mentioned above. These outliers are due to the bright clumps in the nearby galaxy fields, which are more extended than the point sources. The estimated mean FWHM of the PSFs for all UVIT filters is mentioned in \autoref{tab:mag-lim-fit}. We found that the average FWHM of every UVIT filter was less than $1.5''$, as mentioned in \citet{2017tandon}. SE could not find bright sources with the parameters given for 16 fields in various UVIT filters. The PSFs for such images are not included in the catalog. 

\subsection{Source Detection} \label{sec:se-param}

We run SE on science-ready, non-normalized images of all the UVIT filters to detect the sources with a minimum number of 9 pixels (similar to the circular area of PSF FWHM of UVIT) and flux level greater than 3$\sigma$, where $\sigma$ is the standard deviation of the background estimated in \autoref{sec:sky_bkg}. We employ a 34-pixel mesh size for source detection and our calculated mean sky background as the background. The latest zero-point magnitudes of the filters used are listed in \autoref{tab:filter}. We convolve the image using a Gaussian filter of box size $5\times5$ pixels and FWHM of 3.0 pixels. The default values of the deblending subthreshold and minimum contrast parameter in SE are 32 and 0.005, respectively. The UVIT pixel scale, 0.417$''$, is used. The rest of the parameters are kept unchanged from the default SE file. We generate a source catalog using the parameters mentioned above, along with a table with information about the various parameters estimated by SE, such as the background mean, rms, the threshold, and the number of sources detected. The details of the columns included in the catalog are given in \autoref{sec:appendix}.

Generally, the periphery of a UVIT FoV does not get total exposure \citep{ccdlab2017}. Typically, we get 100\% exposure up to a radius of $\approx 12.5'$, and beyond this, the effective exposure value decreases. Hence, there is a high probability of detecting spurious sources with low SNR close the edge of the FOV. To remove these spurious sources, we overplot the sources detected in our catalog on the exposure map (obtained from {\sc ccdlab}) and find the normalized exposure value at the source position. We retain the sources with a normalized exposure value $\ge$ 0.9 and remove the remaining sources from UVIT DR1.
 
\subsection{Photometry} \label{sec:phot}
    
After identifying the coordinates of the sources in the UVIT images, we employ several techniques to perform various kinds of photometry using SE/IRAF, as described below. These methods incorporate the fluxes from both the point and extended sources observed by UVIT. 

\begin{itemize}
   
    \item \textbf{Aperture magnitude}: We provide fixed circular aperture magnitudes within a radius of 3 pixels (1.25$''$; equivalent to the PSF FHWM of UVIT), 12 pixels (5.00$''$; equivalent to {\em GALEX} FWHM), and 15 pixels (6.26$''$; equivalent to aperture radius used for {\em GALEX} AIS catalog). These magnitudes are labeled as MAG\_APER\_X where X is the aperture size (in pixels), in our catalog. The aperture photometry will be useful in analyses involving point sources. However, aperture magnitudes are not available for extended sources i.e., elongated sources having FWHM approximately larger than 3 pixels.
   
    \item \textbf{Isophotal Magnitude}: SE selects all the continuous pixels above the threshold value. It then computes the $\rm 1^{st}$ and $\rm 2^{nd}$ order moments within this region and finds the semi-major (A) and semi-minor (B) axes of an ellipse around the source. The angle between A and the horizontal axis is given by an angle $\theta$. The magnitude within the enclosed region is given by MAG\_ISO. Isophotal magnitude is the most user-independent magnitude and only depends on the detection threshold and the background. It is useful to extract the flux of any halo around the source towards the fainter end.
    
    \item \textbf{Kron Magnitude}: We use Kron photometry \citep{Kron1980} for the extended sources. SE uses the A, B, and $\theta$ parameters derived previously and defines an elliptical aperture having the radii equal to 6 times the parameters A and B. It then computes the first-order moment ($r_{kron}$) and measures the flux within an aperture defined by the kron factor times $r_{kron}$. This aperture is given as Kron radius. We use the kron factor to be 2.5, which encloses 94\% of the total flux of the source \citep{sextractor}. Kron magnitude is given in the catalog by the keyword MAG\_AUTO. Kron magnitudes are particularly useful for extracting the fluxes from extended sources such as clumps in the nearby galaxies, gaseous regions of the nebulae, and other galaxy-like sources. We emphasize here that Kron magnitudes can be used for both point and extended sources. 
    
    \item \textbf{PSF magnitude}: We use the coordinates from the source catalog obtained from SE and the DAOPHOT package in IRAF to measure the PSF. We take the mean PSF FWHM as our aperture radius and our estimated $\sigma$ as the background of the image. The point sources selected to model the PSF profiles of the image were brighter than 20.5 mag with elongation $\le$ 3.0. We removed all the sources that had at least one source within 100 pixels to choose isolated sources. If not enough sources satisfy the criteria, we discard the method, and the catalog consists of only the magnitudes provided by SE. We run the ALLSTAR routine in the DAOPHOT package to apply the model PSF to all the sources and compute their PSF magnitude. We perform aperture correction to PSF magnitudes using the curve of growth derived from the PSF model for fainter sources. The method is described in detail in \cite{stetsoncog}. We use saturation corrections for our PSF magnitudes, as mentioned in \citep{2017tandon}. The aperture corrected PSF magnitudes are given in the catalog by the keyword IRAF\_MAG\_PSF\_COR. These magnitudes are more robust for point sources. However, PSF magnitudes are not available for all the sources.
 
\end{itemize}

\section{Catalog Validation}\label{sec:validate}

In the previous sections, we described the procedure we adopted to prepare UVIT DR1 catalog. While the procedure results in identifying a large set of point and extended sources in the catalog, it may contain various artifacts arising from the presence of saturated stars, UV-bright galaxies, dense clusters, etc. We discuss the measures adopted to remove artifacts from UVIT DR1 catalog in the following subsections. Further, to check the validity of sources present in the catalog, we cross-match non-crowded sources (see \autoref{sec:non-crowded}) with {\em GALEX} catalog and compare the FUV/NUV magnitudes. We also search for optical counterparts of our sources in {\em Gaia} DR3 catalog and deduce the fraction of spurious UV sources in our catalog.

\subsection{Deblending}

 We work with several UVIT observations of the globular clusters to verify the efficiency of the source deblending parameters mentioned in \autoref {sec:se-param}. We plot the extracted sources on various globular clusters where the source density is large compared to non-crowded fields. In globular clusters, SE extracts distinct sources in the outer regions where the sources are further than $\sim$1.5$''$. However, in the inner regions of some of the globular clusters, the source density is so high that the central region appears as a clump, and nearby sources are not well-resolved by SE. SE estimates the flux within the central regions as one or more elliptical kron-like apertures. Hence, these sources are removed from the main UVIT DR1 catalog.

\subsection{Artifacts}

\begin{figure*}
    \centering
    \includegraphics[width=0.31\textwidth]{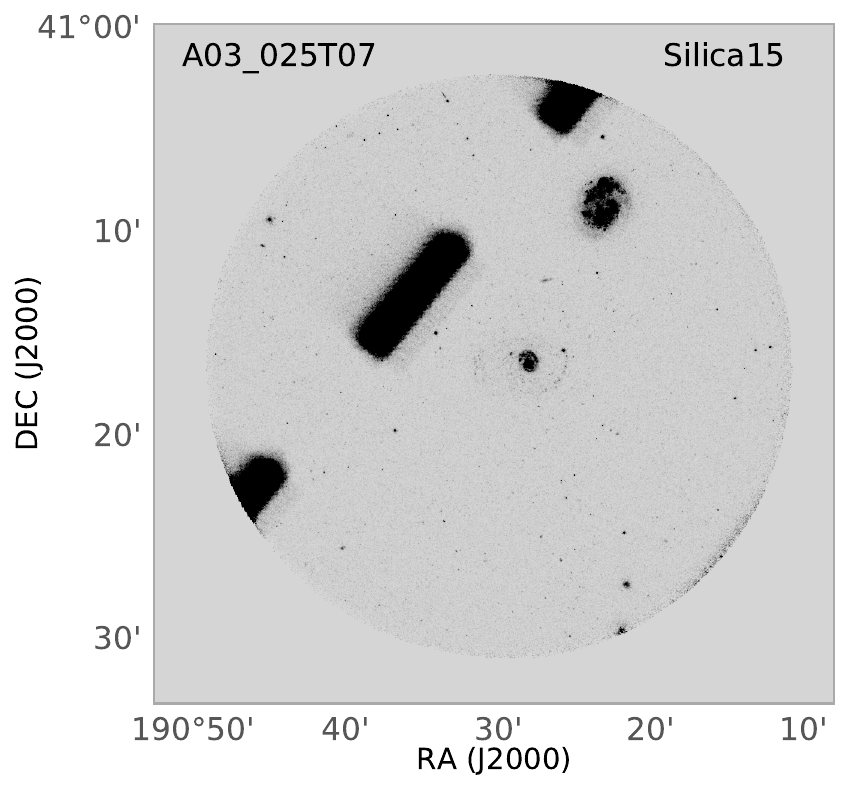}
    \includegraphics[width=0.31\textwidth]{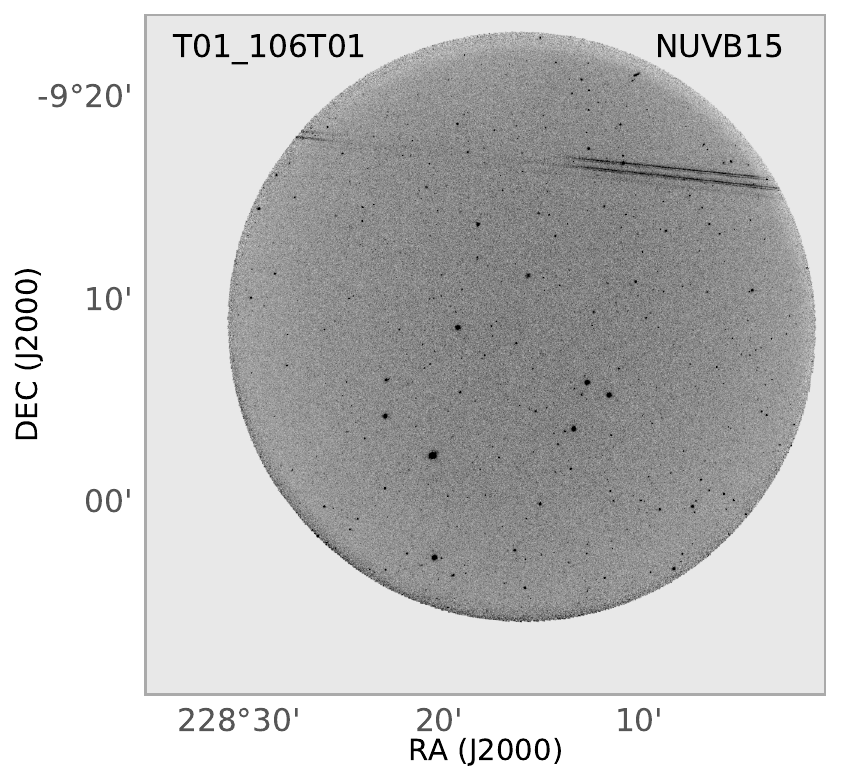}
        \includegraphics[width=0.31\textwidth]{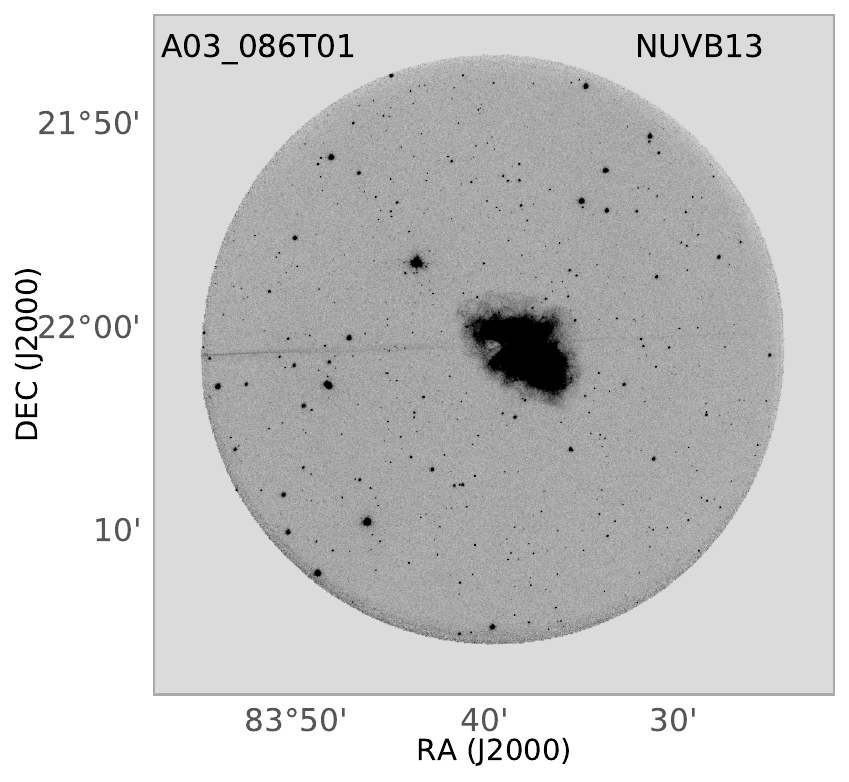}

    \caption{Example of various artifacts observed in UVIT images. Left: Artifacts caused due to large satellite trails. Middle: Artifacts caused due to fast moving satellites. Right: Artifacts caused due to bright galaxy.}

    \label{fig:artifacts}
\end{figure*}	

 As discussed earlier, we encountered various artifacts while constructing our catalog. The diffraction spikes around a saturated star in the FoV are one of the causes of artifacts in the catalog. We ensure we do not include such non-astrophysical objects in our catalog by varying the detection threshold at 3$\sigma$, 2$\sigma$, and 1.5$\sigma$. While SE detects such artifacts at 2$\sigma$ and 1.5$\sigma$, these artifacts are not detected as individual sources at the 3$\sigma$ detection threshold. We verify the same for a few bright stars where diffraction spikes are observed. Hence, this threshold limit is used to extract sources in UVIT DR1. 

 The other artifacts include spurious sources detected due to the reflections of the mirror at the edge of the FoV. We remove these artifacts using the exposure map generated by {\sc ccdlab}, and the procedure is described in detail in \autoref{sec:source_detection}. We also observed patchy regions and one or more bright lines on the FoV of a few UVIT images due to numerous satellite trails in the sky (shown in the left and middle panel of \autoref{fig:artifacts}, respectively). Similar bright lines are also created in the images due to UV-bright stars or galaxies (shown in the right panel of \autoref{fig:artifacts}). During source extraction, SE detects sources along these trails. We have removed the sources caused by these artifacts from the catalog. The catalog, thus obtained, might contain other artifacts, but the fraction of such sources is low.

\subsection{Comparison with GALEX survey} \label{sec:xmatch-galex}

The {\it GALEX} sky survey has covered 70\% of the sky in FUV (1344–1786 \AA, $\lambda_{\mathrm eff}$=1528 \AA) and NUV (1771–2831 \AA, $\lambda_{\mathrm eff}$ = 2310 \AA), and its photometric catalog contains $\sim$ 500 million sources \citep{Bianchi2017}. The {\it GALEX} surveys cover almost 22,125 square degrees of the sky, with a typical depth of 19.9/20.8 mag in FUV/NUV and a spatial resolution of 4.2$''$/5.3$''$ (FUV/NUV). But UVIT has a much better spatial resolution of 1.5$''$ and typical depth in various filters, as mentioned in \autoref{tab:mag-lim-fit}. We have cross-matched UVIT DR1 with the GALEX catalog using a cross-matching radius of 3$''$ for a comparative analysis.

We found that more than 90\% of the matched sources are single-matched, while the rest are multiple-matched, i.e., more than one UVIT counterpart for a single source in the {\em GALEX} catalog, and this is due to the higher spatial resolution of UVIT. For some of the {\it GALEX} sources ($\sim$ 10\%), we do not find a UVIT counterpart within the match radius. These sources lie either at the edges of the UVIT FoV, where the normalized exposure is less than 90\%, or the exposure time of the corresponding UVIT observation is less than that of {\em GALEX}. Although we have cross-matched the UVIT sources detected in different filters with {\it GALEX} sources, we have not included the cross-matched catalog in this paper due to the less sky coverage of UVIT as compared to {\it GALEX}.

We determine the magnitude difference between {\em GALEX} and UVIT Kron magnitudes for the single matched sources using a cross-matching radius of $3''$. The magnitude difference is $\le 1$ for the sources lying at the brighter end in FUV filters. However, for NUV filters the magnitude difference is on an average greater than 1 even at the brighter side of magnitude scale. Since the limiting magnitude of the UVIT is greater than {\em GALEX}, UVIT detects the sources at much fainter magnitudes. These sources have a magnitude difference greater than 1 mag in both these filters.  This is a useful sanity check to show, among other things, that the cross-matched sources are identified correctly. The wavelength ranges covered by the UVIT and GALEX filters are slightly different, hence it is unlikely that the magnitudes would be the same.

\subsection{Comparison with Gaia DR3}
\label{sec:xmatch-gaia}

\cite{gaiadr1} recently had its third data release, {\it Gaia} DR3 \citep{gaiadr3}, which includes astrometry and photometry for 1.8 billion objects covering almost the entire sky. The catalog contains photometry in three broad-band filters, $G$, $G_{BP}$, and $G_{RP}$, radial velocities, and parallaxes of the sources. {\it Gaia} DR3 has a resolution of 1.5$''$ beyond which the completeness falls rapidly \citep{gaia-resolution}. We have downloaded {\it Gaia} DR3 catalogs from {\it Gaia} Archive using the central coordinates of our FoVs for an area of radius 15$'$ to get Gaia sources covering the entire FoV of UVIT. Since the resolution of {\it Gaia} DR3 is similar to UVIT, we use a cross-matching radius of 1.5$''$ to cross-match our catalog with the {\it Gaia} DR3 catalog. We found most of the sources are single-matched (more than 90\%).

The slightly extended (size less than 5$''$) sources detected as single sources in UVIT have multiple counterparts in Gaia. We also did not find Gaia counterparts for some of the bright UV sources of UVIT. These sources emit flux predominantly in the UV waveband, and their magnitude ranges between 14.5 and 22.5 mag. More than 99\% of the matched sources have G mag, while 86$-$94\% of the sources have $G_{BP}$ or $G_{RP}$ mag. Of the total matched sources, around 70$-$90\% sources in FUV have parallax values in {\it Gaia} DR3, while 86$-$92\% sources in NUV have parallax values available in {\it Gaia} DR3.

\subsubsection{Spurious Sources}

We followed the procedure mentioned in \citet{2011bianchi} to find the spurious matches for sources observed in the non-crowded field sources in UVIT DR1 and {\em Gaia} catalog. To do this, we generated a random coordinate at an offset of 0.5$'$ from the matched UVIT$-${\em Gaia} catalog and cross-matched these offset coordinates to find their counterparts in {\em Gaia} DR3 catalog using a radius of 1.5$'$. We found spurious matches for $\sim$0.1$-$0.15\%  of the total matched sources in NUV filters. These matches may include positional coincidences. However, we did not find any spurious matches in FUV filters. We have applied the same procedure for the whole catalog and found the spurious matches for $\sim$1$-$2\% of the total matched sources in NUV filters and $\sim$1$-$1.5\% in FUV filters.

\subsection{Comparison with M31 galaxy fields from \citet{Leahy2020}} \label{sec:m31}

\begin{figure*}

    \centering
    \includegraphics[width=0.48\textwidth]{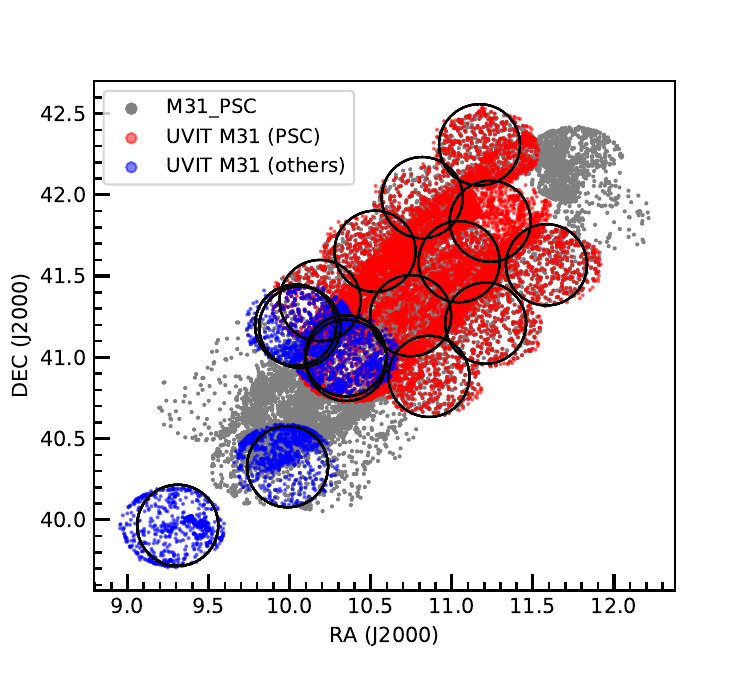}
    \includegraphics[width=0.48\textwidth]{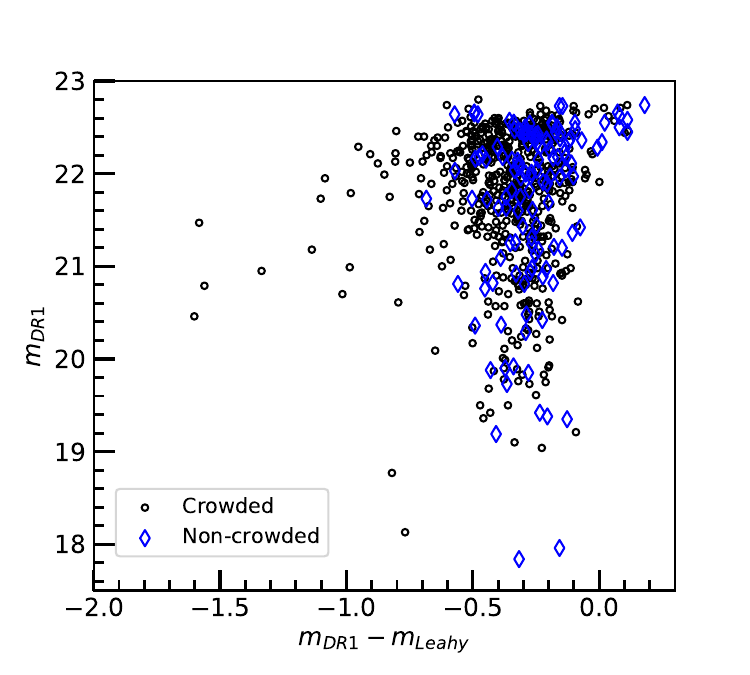}
    \caption{The area covered by M31 FoVs observed by UVIT in various UVIT filters and the black solid circles show the FoV of the image. The red dots are the UV sources in UVIT DR1 overlapped with the UV source catalog published by \citet{Leahy2020} (represented by grey dots). The blue dots are the sources in UVIT DR1, which are observed by other PIs. In the right panel, we show the magnitude difference between UVIT DR1 and UV source catalog published in \cite{Leahy2020} for a single M31 field (A04\_022T07). The magnitude difference increases in the crowded regions.}
		\label{fig:m31-area}
\end{figure*}

UVIT has observed 16 FoVs of the M31 galaxy over two years in both FUV and NUV filters. The sources extracted from all M31 observations in our catalog are shown by red and blue solid points in \autoref{fig:m31-area}. The UVIT pointings of M31 included in UVIT DR1 overlap with 11 fields published by \citet{Leahy2020}. Actually, the UV sources published by \citet{Leahy2020} include a total of 19 fields of UVIT observations, of which only 11 fields were observed during 2016$-$2017. The grey dots in \autoref{fig:m31-area} show the total sky area of M31 covered by \citet{Leahy2020} while the red dots show the 11 fields of M31 included in UVIT DR1. The blue solid points are the UV sources observed in M31 by other PIs with PIDs A02\_197 and A03\_044. The number of M31 fields observed in different UVIT filters is listed in \autoref{tab:fieldtype}. The UV sources of M31 are flagged with flag {\it ID} = 1 in the catalog. 

The M31 point source catalog \citep{Leahy2020} used the {\sc ccdlab} in-built point source extraction algorithm to detect the point sources above 3$\sigma$ of the local background. The sources were fitted using elliptical Gaussian, and magnitudes were corrected using the COG analysis in {\sc ccdlab}. The catalog contains the point sources observed in 5 different UVIT filters: CaF2, Sapphire, Silica, NUVB15, and NUVN2. The catalog gives the PSF magnitudes of these sources. In UVIT DR1, we include the fixed aperture magnitudes of sources in five sizes (see \autoref{sec:phot}), the PSF magnitudes of the point sources, the Kron, isophotal, and Petrosian magnitudes of the sources. We also report various morphological parameters, such as A, B, elongation, and ellipticity, of the clumps detected in the M31 galaxy.

We cross-matched the UV sources obtained in UVIT DR1 with the UV sources of M31 published in \citet{Leahy2020}, and the number of sources matched is given in \autoref{tab:m31-xmatch}. The point sources in the non-crowded regions in UVIT DR1 match well, but discrepancies occur mainly in FUV observation of the clumps in spiral arms. The different methods used to detect the sources and calculate the magnitudes could be a possible reason for this discrepancy. The right panel in \autoref{fig:m31-area} shows the magnitude difference between UVIT DR1 and M31 point source catalog in crowded and non-crowded regions of a single field of M31 (A04\_022T07). The magnitude difference is $\sim$0.5 mag for the sources in non-crowded regions (blue diamonds), but it increases up to $\sim$2 mag in crowded regions. The match percentage of Leahy sources is 47\% in CaF2 and NUVB15 filters and 59\% in Silica and NUVN2 filters. The unmatched sources lie in the crowded regions, which could be due to the deblending effect. UVIT DR1 has detected a whole clump in the crowded regions, whereas, in the M31 point source catalog, multiple sources are detected.

\begin{table}
	    \centering
	    \caption{Number of point sources obtained from our catalog after cross-matching with M31 PSC. $N_{sCat}$ and $N_{sLeahy}$ are the number of sources detected in UVIT DR1 and M31 catalog published by \citet{Leahy2021}. x\_M31\_Leahy column gives the number of sources matched using a crossmatching radius of 1$''$, 1.5$''$, and 2$''$.}
	   
	    \begin{tabular}{cccccc}
	    \hline
	          \multirow{2}{*}{Filters} & \multirow{2}{*}{$N_{sCat}$} & \multirow{2}{*}{$N_{sLeahy}$} & \multicolumn{3}{c}{x\_M31\_Leahy} \\
	         (X-match radius) & & & (1$''$) & (1.5$''$) & (2$''$) \\ 
           
	         \hline
	         CaF2 & 14786 & 20925 & 9824  & 11074 & 11522\\
	         Silica & 6269 & 6577 & 3910 & 4395 & 4594\\
              NUVB15 & 9569 & 15428 & 7328 & 8061 & 8176 \\
              NUVN2 & 6991 & 8532 & 5009 & 5157 & 5288 \\
	         \hline
	    \end{tabular}
	    
	    \label{tab:m31-xmatch}
	\end{table}

\section{Details of the UVIT source catalog} \label{sec:cat_detail}

We have constructed a UV source catalog of all the 291 FoVs observed by UVIT during 2016$-$2017. These FoVs cover an area of $\sim$58.2 deg$^2$. The number of fields observed in various UVIT filters and the total number of sources ($N_{s}$) detected in them are provided in \autoref{tab:source_numbers}. 

We provide the catalog, UVIT DR1, of all the UV sources observed by UVIT during 2016$-$2017 in an electronic table that will be available online with the final version. In UVIT DR1, we provide information about the source position, their shape (semi-major axis, semi-minor axis, position angle), and UV fluxes estimated in nine UVIT filters by applying various photometric methods, i.e., aperture, Kron, Petrosian, and PSF photometry (described in \autoref{sec:phot}). The description of all columns in the catalog is given in \autoref{sec:appendix}.

To identify the multi-epoch observations with different PID and TID values, we use an internal cross-matching radius equivalent to the effective PSF FWHM of the filters, as mentioned in \autoref{tab:mag-lim-fit}. The sources that have unique observations have $N_{epoch}$ {\it ID} `0\_0' in the catalog, while multi-epoch observations have $N_{epoch}$ {\it ID} `i\_j'. All the sources observed twice or more will have the same `i' values, and the `j' denotes the number of occurrences of that source. The number of multi-epoch sources in each filter is given by the row N$_{multi}$ in \autoref{tab:source_numbers}.

The magnitudes of the UV sources in UVIT DR1 are in the range of 11 to 26 mag. The error in magnitudes for most of the sources is $\le$0.5 mag except for a few sources ($\sim$200 sources in all the filters). The exposure times of these sources are less ($\sim$150 s), which could be the reason for their high error. In the catalog, we have placed a flag of 99 in the magnitude columns wherever SE or IRAF could not measure magnitudes. 

\begin{table*}
    \centering
    \caption{Total number of FoVs ($N_{F}$) and number of sources ($N_{s}$) observed in each filter of UVIT during the year 2016-2017. 
    }
    \label{tab:source_numbers}
    %\adjustbox{max width=\textwidth}{
    \begin{tabular}{ccccccccccccccccccc} % four columns, alignment for each
    \hline
    
 Year & \multicolumn{2}{c}{CaF2} &
  \multicolumn{2}{c}{BaF2} &
  \multicolumn{2}{c}{Sapphire} &
  \multicolumn{2}{c}{Silica} &
  \multicolumn{2}{c}{Silica15} &
  \multicolumn{2}{c}{NUVB15} &
  \multicolumn{2}{c}{NUVB13} &
  \multicolumn{2}{c}{NUVB4} &
  \multicolumn{2}{c}{NUVN2} \\ 
  \hline
  
& $N_{F}$ & $N_{s}$ & $N_{F}$ & $N_{s}$ &$N_{F}$ & $N_{s}$ & $N_{F}$ & $N_{s}$ &  $N_{F}$ & $N_{s}$ & $N_{F}$ & $N_{s}$ & $N_{F}$ & $N_{s}$ & $N_{F}$ & $N_{s}$ & $N_{F}$ & $N_{s}$ \\
\hline
2016 & 34 &	15901 &	35	&	6787 & 49 & 11146 & 26 & 3732 & 20	& 17022	& 36 & 12144 &	45	& 26934	& 41 & 23110 &	46	& 19330\\
\hline
2017 & 61 & 35881 &	71 & 14732 & 60 & 14967 & 49 & 17609 &	56 & 75702 & 44 & 15616 & 70 & 36964 & 65 & 59171 & 54 & 22962\\
\hline
Total & 95 & 51782 & 106 & 21519 & 109 & 26113 & 75 & 21341 & 76 & 92724 & 80 & 27760 & 115 & 63898 & 106 & 82281 & 100 &	42292 \\
\hline
 N$_{multi}$ & 22 & 19857 & 30 & 2533 & 25 & 8692 & 20 & 1334 & 11 & 11306 & 14 & 7632 & 21 & 7910 & 15 & 7786 & 36 & 8923 \\  
\hline

\end{tabular}
%}
\end{table*}

\begin{table*}
    \centering
    \caption{Various types of fields observed by different filters of UVIT telescope.}
    \label{tab:fieldtype}
    \begin{tabular}{ccccccccccccc} % four columns, alignment for each
    \hline
    \hline
    \multirow{2}{*}{Field-type} & \multirow{2}{*}{Flag} & 2016 & 2017 & \multirow{2}{*}{CaF2} & \multirow{2}{*}{BaF2} & \multirow{2}{*}{Sapphire} & \multirow{2}{*}{Silica} & \multirow{2}{*}{Silica15} & \multirow{2}{*}{NUVB15} & \multirow{2}{*}{NUVB13} & \multirow{2}{*}{NUVB4} & \multirow{2}{*}{NUVN2}  \\
 & & ($N_{F})$ & ($N_{F}$) \\
    \hline

M31 Galaxy & 1 &  5 & 11 & 10 & 4 & 5 & 12 & - & 13 & 5 & 5 & 10\\
Globular Clusters & 2  & 6 & 8 & 5 & 5 & 9 & 3 & 2 & 4 & 6 & 7 & 3  \\
Open Clusters & 3 & 3 & 7 & 8 & 2 & 6 & 1 & 1 & 2 & 2 & 1 & 6\\
Gaseous nebulae & 4 & 7 & 4 & 1 & 5 & 10 & 4 & - & 2 & 2 & 1 & 9 \\
Planetary Nebulae & 5 & 6 & 7 & - & 3 & 10 & 10 & - & 3 & 9 & 9 & 3\\
Magellanic Clouds & 6 & 4 & 10 & 9 & 3 & 4 & 4 & 6 & 4 & 2 & 5 & 3\\
Nearby Galaxies & 7 & 27 & 72 & 35 & 37 & 22 & 23 & 34 & 26 & 37 & 42 & 22\\ 
Non-crowded Fields & 0 & 77 & 139 & 60 & 85 & 80 & 54 & 67 & 56 & 100 & 84 & 76\\

\hline
		
\end{tabular}
\end{table*}
\subsection{Non-crowded Fields}
\label{sec:non-crowded}
\begin{figure*}
    \centering
    \includegraphics[width=0.48\textwidth]{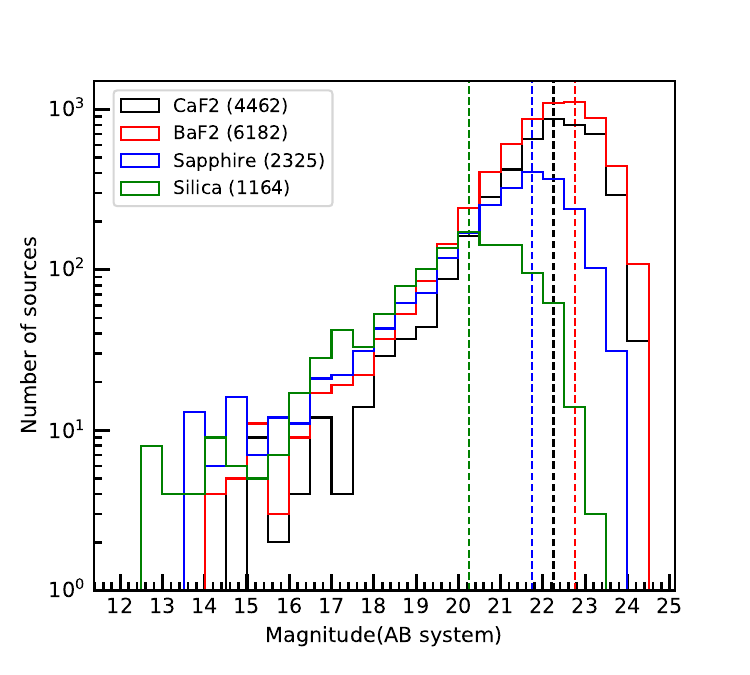}
    \includegraphics[width=0.48\textwidth]{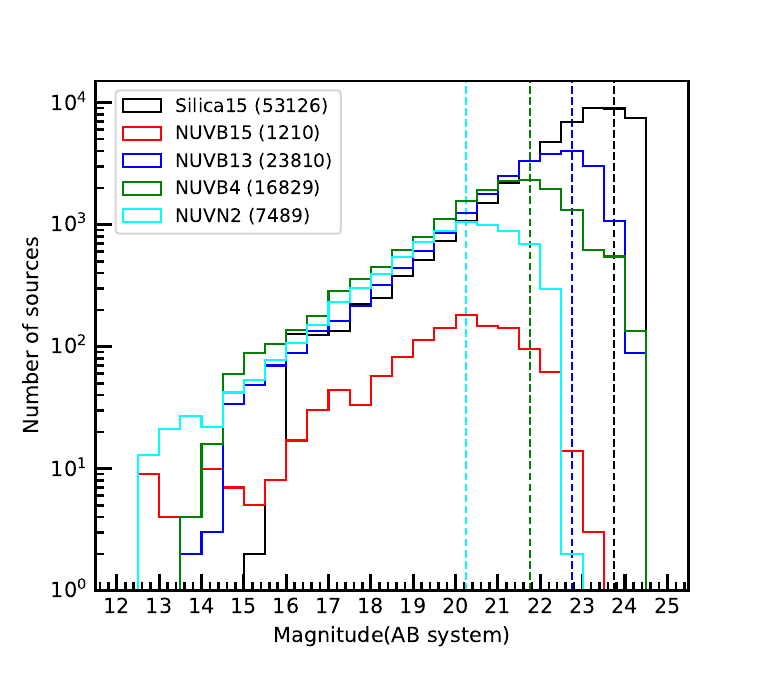}
    \caption{Magnitude histograms in various FUV (left panel) and NUV (right panel) filters of UVIT. {\bf The dashed lines represent the peak magnitudes of the respective filters, which determines the magnitudes at which most sources are detected}. The filter names, along with the number of sources detected, are mentioned in the legend.} 
    \label{fig:count-hist-fields}
\end{figure*}	

The sources other than the large angular-sized objects, such as star clusters, nebulae, and nearby galaxies, are categorized as non-crowded field sources in our catalog. These sources are flagged with Flag {\it ID} $=$ 0. 

\autoref{fig:count-hist-fields} shows the magnitude histogram of UVIT non-crowded field sources in various FUV and NUV filters. The dotted lines in \autoref{fig:count-hist-fields} represent the magnitude at which the magnitude histogram peaks. This peak value approximately determines the depth of the detection of the sources in the catalog. We detect $\sim$68\% sources up to 22.75 mag in CaF2 and 23.75 mag in Silica15 filters. 

The underlying objects with Flag {\it ID} $=$ 0 could be stars, galaxies, or distant quasars. One would require spectroscopic classification to separate them into different classes. We devise a method to classify point and extended sources using FUV and NUV magnitudes in \autoref{sec:star-galaxy}, which may be used in the unavailability of spectroscopic classification. 

\subsubsection{Star-Galaxy Separation}
\label{sec:star-galaxy}
\begin{figure*}
    \centering
    \includegraphics[width=0.48\textwidth]{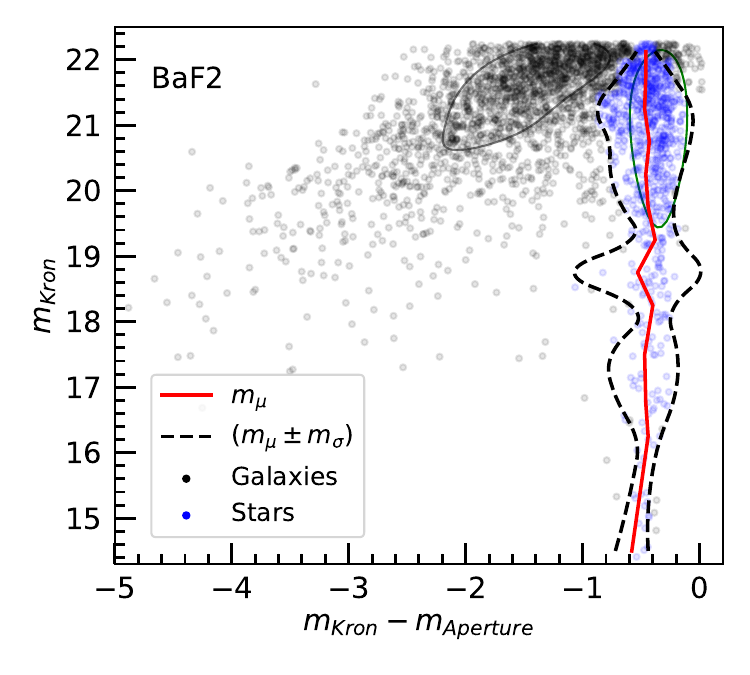}
    \includegraphics[width=0.48\textwidth]{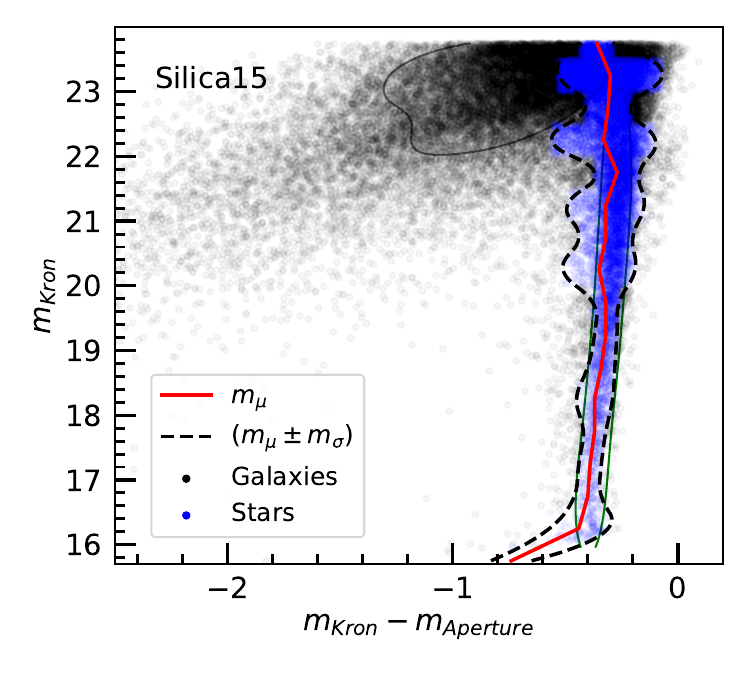}

    \caption{We cross-match the non-crowded UVIT field sources in our catalog with SIMBAD. The difference in their aperture magnitude (aperture radius$=$3 pixels) and Kron magnitude are shown in the left panel. The gray and green contours represent the confirmed stars and galaxies from the SIMBAD catalog. We calculate the $m_{\mu}$ and $m_{\sigma}$ for (m$_{Kron}$ $-$ m$_{Aperture}$) in each magnitude bin of bin width $=$ 0.5 mag. The red solid line shows the $m_{\mu}$ line, and the black dashed line shows the $m_{\mu}\pm m_{\sigma}$ lines. The sources between these lines are classified as probable point sources, while the rest are probable extended sources.}    
 
    \label{fig:field-point-extended}
\end{figure*}	

Several methods based on photometric measurements have proven to effectively separate point and extended sources in cases where spectroscopic observations are unavailable. The stars and galaxies are shown to follow distinct distributions on various UV-optical or UV-IR color planes \citep{Bianchi2007,2014ananta,Bianchi2020}. A similar classification is also possible using magnitude measurements from a single broadband filter. The difference in PSF and total magnitude for optical observation of sources is a widely used parameter to separate stars and galaxies \citep{2002Strauss, 2010Baldry}. We were not in capacity to compute colors for all sources in our catalog due to a lack of optical and IR counterparts for a sizable sample. Hence, we use the FUV/ NUV magnitudes for the segregation.

We cross-match the non-crowded field sources in our catalog (as described in \autoref{sec:non-crowded}) with SIMBAD to separate the confirmed stars and galaxies from our catalog. To classify the probable point sources, we bin the m$_{\rm Kron}$ of confirmed SIMBAD stars of bin width 0.5 mag and find the mean (m$_{\mu}$) and standard deviation (m$_{\sigma}$) corresponding to $\rm m_{Kron} - m_{Aperture}$ (aperture radius $=$3 pixels) in each bin. We classify the sources between $m_{\mu}\pm m_{\sigma}$ as probable point sources, and the source type is given in the catalog by column `NS/G'. The NS/G value 0 indicates a probable extended source, and 1 indicates a probable point source. We repeat the analysis for each filter separately. \autoref{fig:field-point-extended} shows the $\rm m_{Kron} - m_{Aperture}$ vs. $\rm m_{Kron} $ plot in BaF2 and Silica15 filters. The blue and black solid circles show the probable point sources and extended sources, respectively. This classification validates up to $\sim$21.5 mag in FUV filters and $\sim$23 mag in NUV filters. Beyond this limit, the deviation increases, and thus, there is a higher percentage of contamination of galaxies in point sources.

Our catalog includes UV sources observed in 16 fields of the M31 galaxy, 14 fields of galactic globular clusters, 10 fields of open clusters, 11 fields of gaseous nebulae, 14 fields of Magellanic clouds, and 8 fields of planetary nebulae. We provide separate flags for the sources observed in such fields for their further analysis. In \autoref{tab:fieldtype}, we have given the flag {\it ID} and the total number of fields observed in different UVIT filters for various sources. Note that we use the same parameters for source detection and photometry throughout the paper to maintain uniformity unless otherwise mentioned. In the following subsections, we describe a few specific categories of sources present in UVIT DR1.

\subsection{Galactic globular clusters} \label{sec:gc}

UVIT has observed 12 galactic globular clusters (GCs) in 14 FoVs during 2016$-$17, with NGC 5466 and NGC 5053 being observed at two epochs each. We use the source extraction procedure described in \autoref{sec:source_detection} to extract the sources from the GCs. The UV sources in globular cluster fields are flagged with Flag {\it ID} = 2. However, the source extraction procedure used in the paper could not resolve the central dense region of the GCs. Hence, we exclude all the sources detected in the central region of GCs within a radius (R) given in \autoref{tab:gc-core-mag}.

We cross-match the resolved UV sources with the globular cluster catalog of {\em Gaia} EDR3 \citep{gaia-gc} using a matching radius of 1.5$''$ and select the cluster members among the detected sources with a membership probability higher than 80\%. We corrected the UV magnitudes of the cluster member sources for extinction using the extinction law provided in \citet{cardelliext}. The extinction-corrected UV magnitudes were scaled to their absolute magnitude using cluster distance taken from \citet{gc-dist}. 

\begin{figure*}
    \centering
        \includegraphics[width=0.3\textwidth]{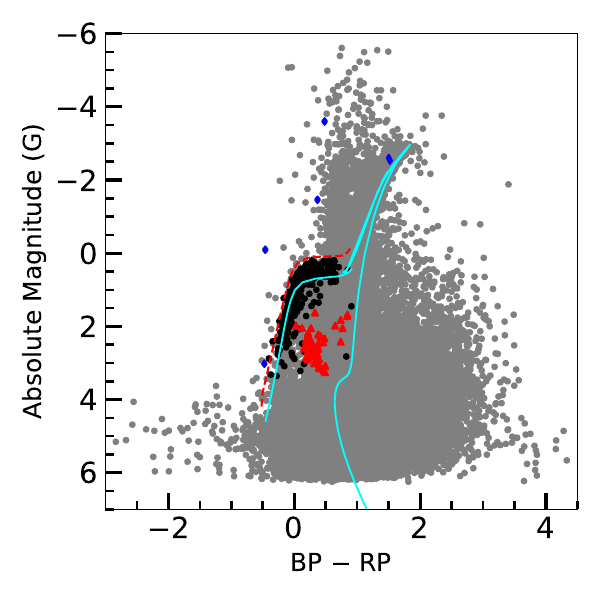}
        \includegraphics[width=0.6\textwidth]{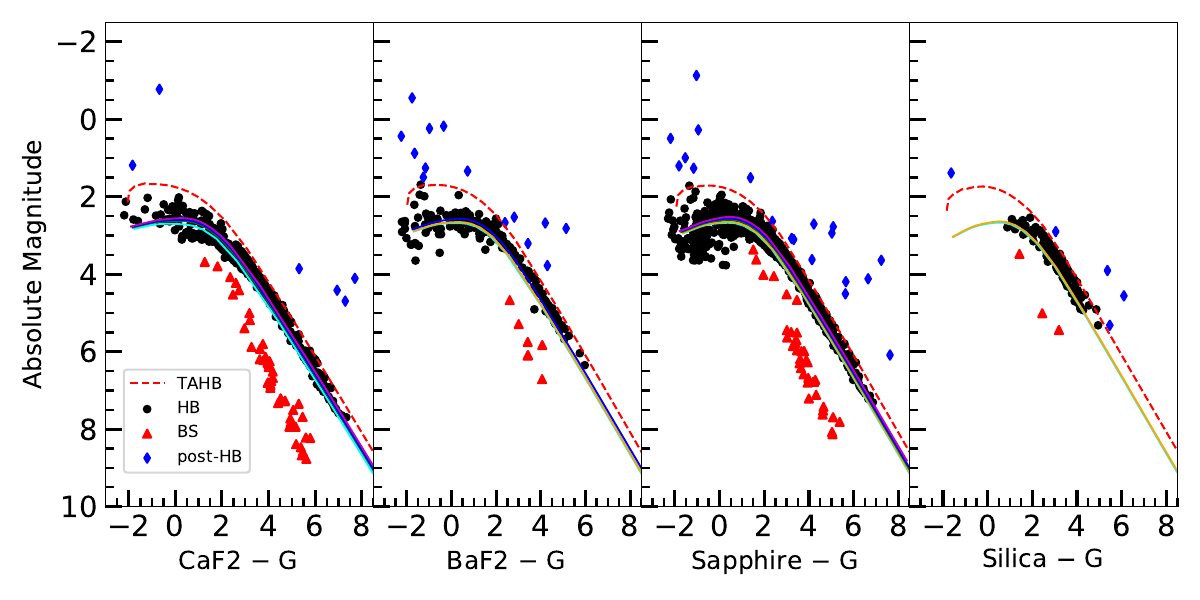}\\

    \includegraphics[width=0.3\textwidth]{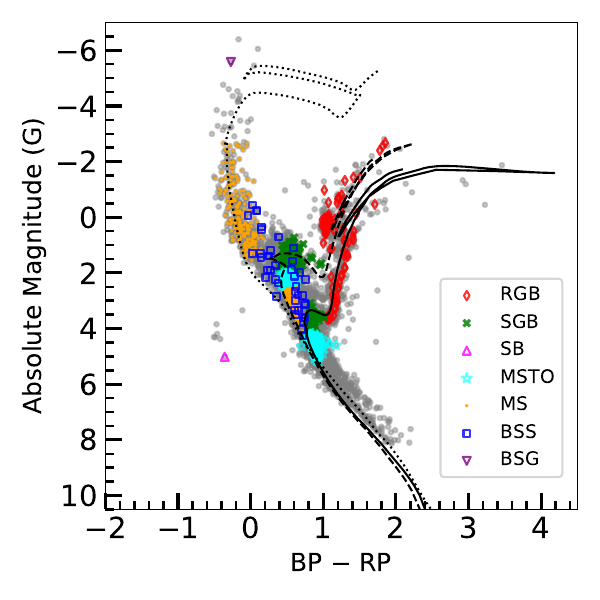}
     \includegraphics[width=0.6\textwidth]{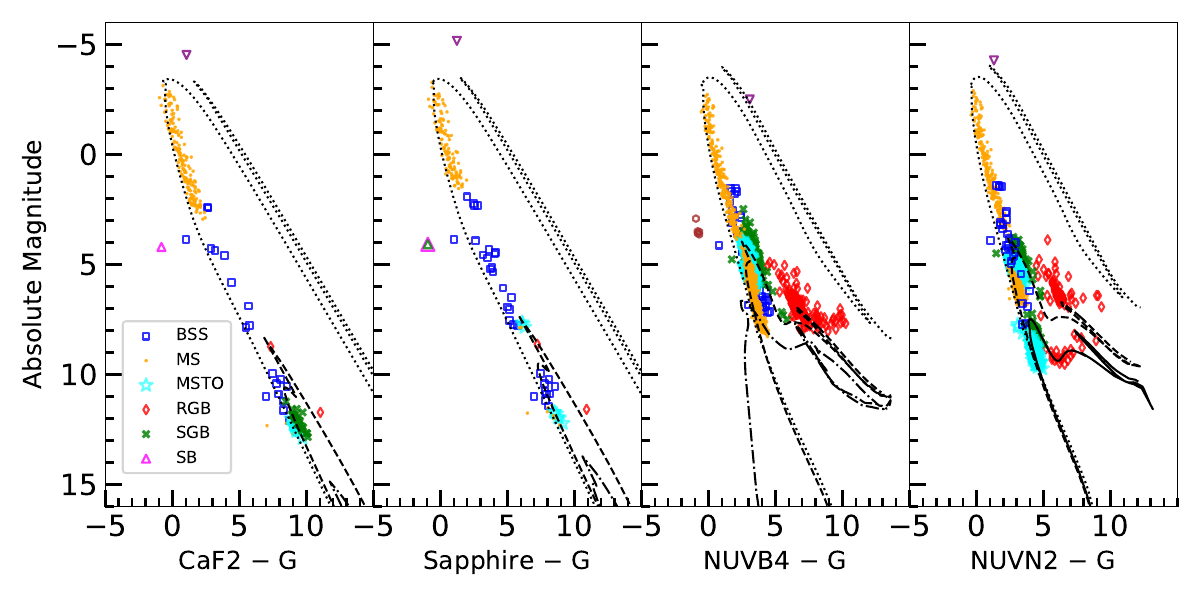}

    \caption{Top left: Optical CMD using {\it Gaia} colors BP $-$ RP vs. G magnitude of all the sources of the 12 globular clusters. The grey dots are the {\it Gaia} sources from \cite{gc-dist}, and the black circles, red triangles, and blue diamonds are the HB, BS, and post-HB stars observed in the CaF2 filter. Top right: The UV-optical CMD plots of CaF2, BaF2, Sapphire, and Silica filters are shown from left to right. We plot the absolute magnitude of the respective filters on the y-axis. The solid lines represent the zero-age HB (ZAHB) loci at [Fe/H]$= -1.20$ (cyan), $-1.30$ (orange), $-1.55$ (green), $-1.70$ (blue), $-1.90$ (magenta), and the red dashed line represents the TAHB line at [Fe/H]$= -2.20$. Bottom left: Optical CMD of all the open clusters observed in UVIT. The gray dots are the cluster members from \cite{oc_cantat}. We have separated the sources observed in the NUVN2 filter into various evolutionary phases: BS (blue boxes), main-sequence (MS; orange dots), main-sequence turn-off (MSTO; cyan stars), RGB (red diamonds), sub-giant (green crosses), spectroscopic binary (magenta upward triangles), blue super-giant (purple downward triangle), and sub-dwarf stars (brown hexagons). Bottom right: The UV$-$optical CMD plots of CaF2, Sapphire, NUVB13, and NUVN2 from left to right. We plot the absolute magnitude of the respective filters on the y-axis. The BaSTI isochrones are plotted at [Fe/H]$=-$1.25 (dotted line), +0.02 (dashed line), +0.03 (dashed dot line), +0.12 (solid line).}
    \label{fig:gc-gaia-hb}
\end{figure*}

We construct FUV-optical CMDs as shown in \autoref{fig:gc-gaia-hb} for all cluster members observed with UVIT FUV filters and {\em Gaia} G-band filter and find that the blue HB (BHB) stars are the most dominant source of UV emission in probed clusters. We plot the zero-age HB (ZAHB) loci from BASTI isochrones\footnote{\href{http://basti-iac.oa-abruzzo.inaf.it/index.html}{http://basti-iac.oa-abruzzo.inaf.it/index.html}} of various metallicities ([Fe/H]) ranging from $-$1.20 to $-$1.90 at the age of 12 Gyr. The isochrones are shown as solid lines in \autoref{fig:gc-gaia-hb}, and the dotted line shows the terminal-age HB (TAHB) loci at metallicity [Fe/H]$ = -$2.20. We find that the metallicity variation has a negligible effect on the FUV$-$ optical colors and FUV magnitudes. The HB stars lie between the ZAHB and TAHB loci. The column `Stype' in the globular cluster catalog indicates the evolutionary phases of the stars. We detect 500 HB stars (`Stype'=hb) in CaF2, 211 in BaF2, 758 in Sapphire, and 134 in Silica filters. The hot post-HB stars lie above the TAHB loci, and the hot BS stars lie below the ZAHB loci in the upper panel of \autoref{fig:gc-gaia-hb}. We find 45 BS stars (`Stype'=bss) in CaF2, 7 in BaF2, 37 in Sapphire, and 3 in Silica. We identify 6 post-HB stars (`Stype'=post-hb) in CaF2, 14 in BaF2, 19 in Sapphire, and 5 in Silica.

\begin{figure*}
    \centering
\includegraphics[width=0.32\textwidth]{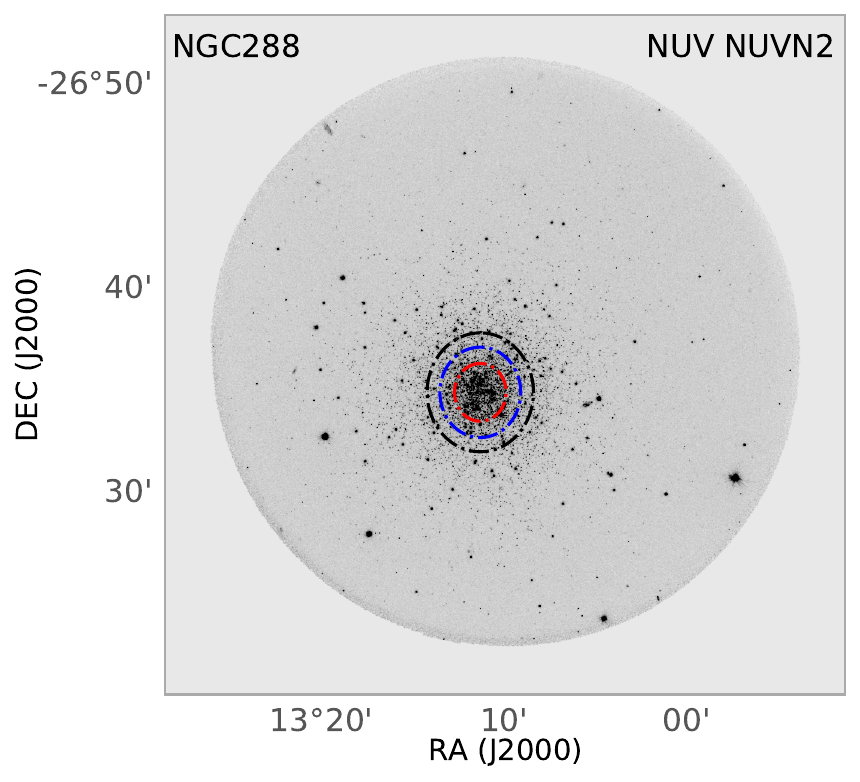}
\includegraphics[width=0.32\textwidth]{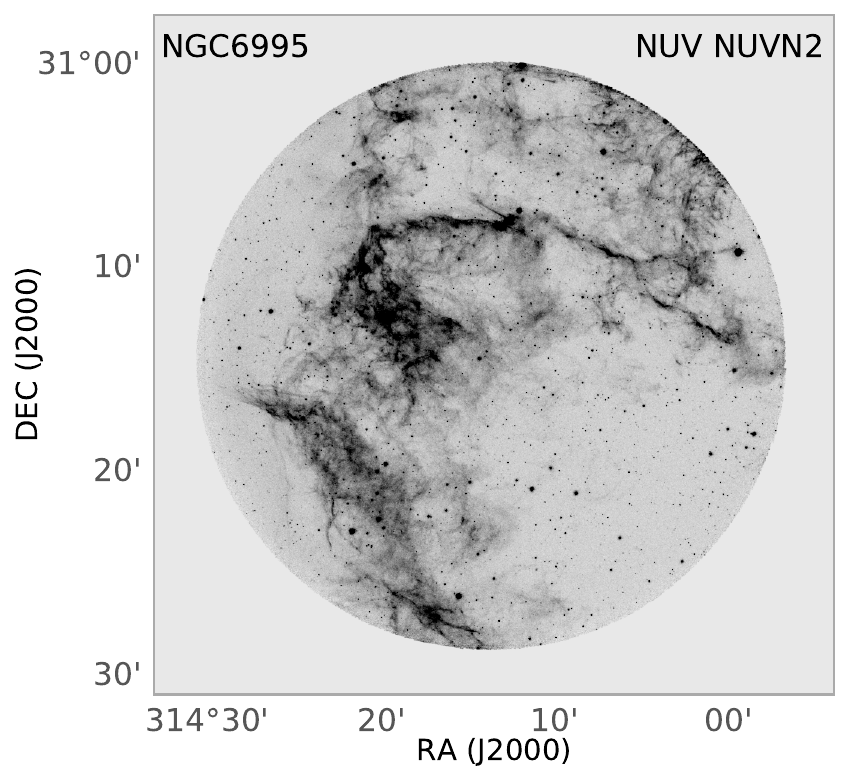}
\includegraphics[width=0.32\textwidth]{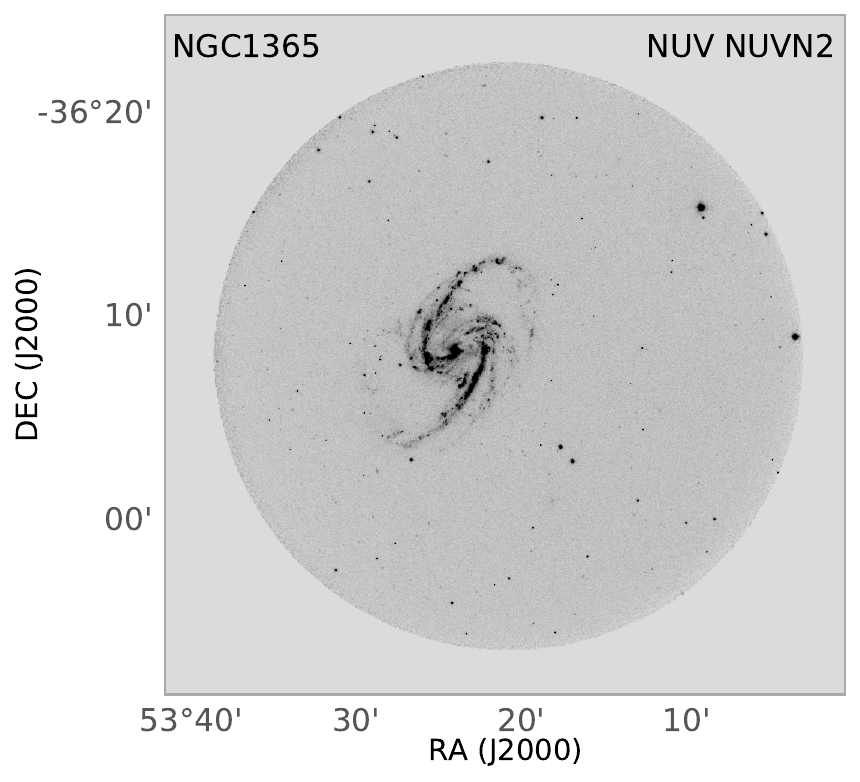}\\
\includegraphics[width=0.32\textwidth]{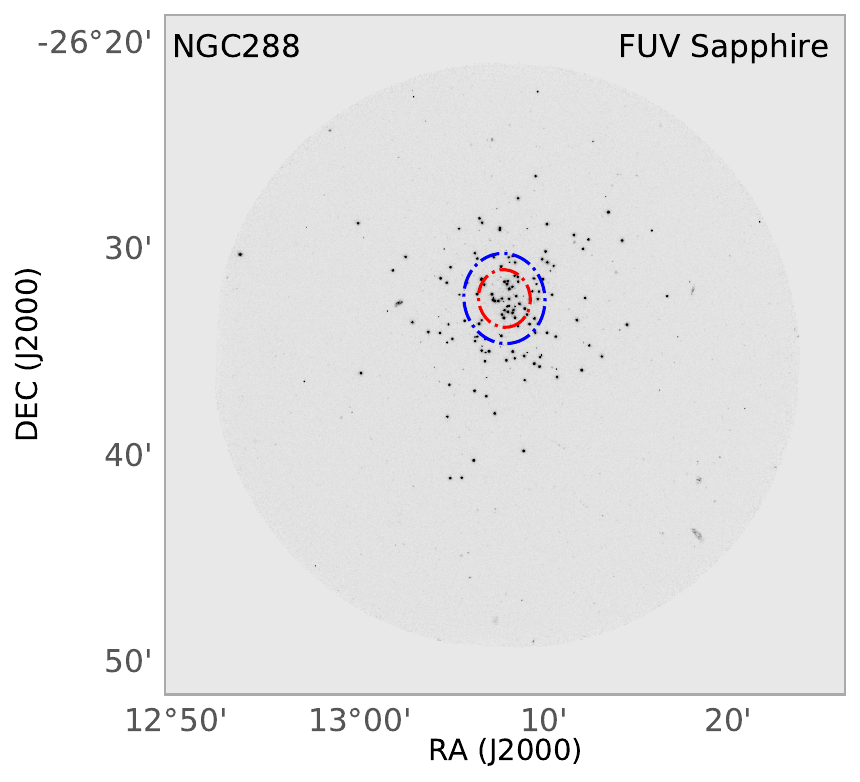}
\includegraphics[width=0.32\textwidth]{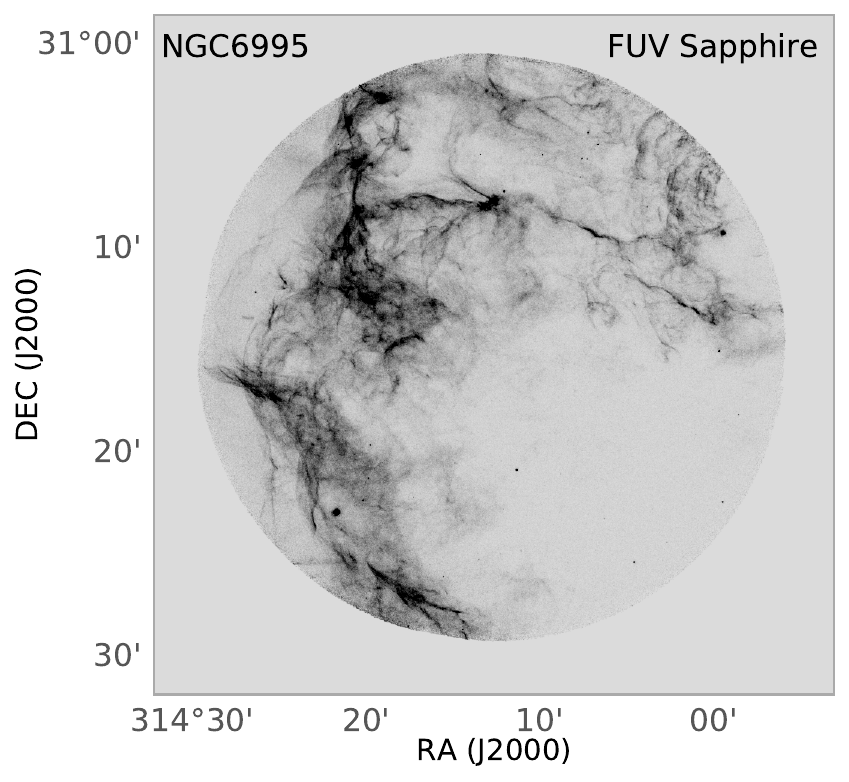}
\includegraphics[width=0.32\textwidth]{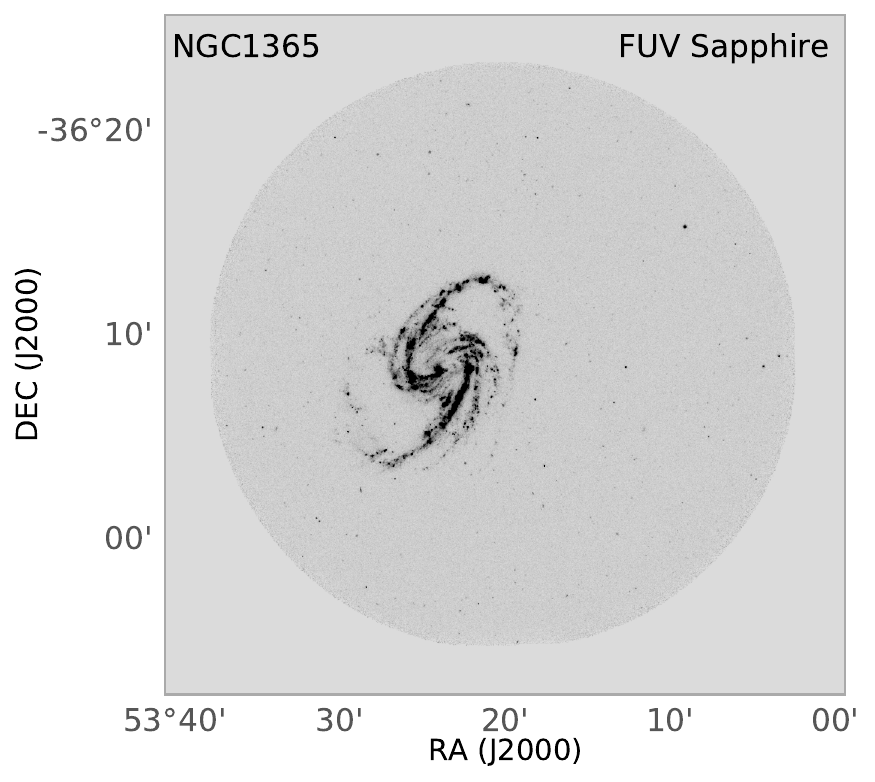}

    \caption{Image of a few astrophysical objects observed by UVIT in FUV and NUV filters. Left: UVIT observation of globular cluster NGC288 (G05\_009T02). The red, blue, and black circles represent the r$_{c}$, r$_{h}$, and R (radius within which the sources are unresolved by the source-extractor), respectively. Middle: Point sources and clumps in gaseous nebula NGC6995 (G06\_060T01). Right: Clumps in the nearby galaxy NGC1365 (A02\_060T01). } 
    \label{fig:sample-field}
\end{figure*}

For the unresolved sources inside the circular region, R, we calculate an integrated aperture magnitude (m$_R$) using the DS9. These magnitudes are included in \autoref{tab:gc-core-mag}. In the table, we also provide the integrated UV core magnitude (m$_{c}$) and half-light magnitude (m$_{h}$) of the observed globular clusters within their core radius (r$_{c}$) and half-light radius (r$_{h}$), respectively. The values of r$_{c}$ and r$_{h}$ are taken from \cite{1996harrisgc}. The left panel of \autoref{fig:sample-field} shows the r$_{c}$ (red circle), r$_{h}$ (blue circle), and R (black circle) in the cluster NGC288 observed in FUV and NUV filters. We see that the sources in the FUV filters are better resolved by UVIT than the NUV filters. For clusters where all the sources are resolved in the core, only m$_{c}$ and m$_{h}$ magnitude values are given.

A detailed analysis of the UV sources identified in the observed globular clusters will be performed in the upcoming papers.

\begin{table}
		\centering
		\caption{Integrated magnitudes of the clusters at different radii. $R$ is the radius within which the sources in the central regions are unresolved by source-extractor, and $m_{R}$ is the magnitude within this radius. m$_{c}$ and m$_{h}$ are the magnitudes within the core radius (r$_{c}$) and half-light radius (r$_{h}$) of the cluster.}
	   	\label{tab:gc-core-mag}
\begin{tabular}{cccccc}
\hline
    Name & Filter &  R & m$_{R}$ & m$_{c}$ & m$_{h}$   \\
    \hline
  \multirow{4}{*}{NGC7492} & BaF2 & - & - & 16.67 & 16.15 \\
  & NUVB15 &  - & - & 15.44 & 14.92 \\
  & NUVB13 & - & - & 15.48 & 15.02 \\
  & NUVB4 & 0.59$'$ & 15.66 & 15.07 & 14.58 \\
  
  \hline
    \multirow{2}{*}{NGC4590}  & NUVB13 & 2.21$'$ & 11.97 & 13.37 & 12.26 \\
 
  & NUVB4 & 3.17$'$ & 11.30 & 12.96 & 11.83 \\ 
  \hline  
  \multirow{3}{*}{NGC1904} & CaF2 & 1.14$'$ & 12.78 & 15.04 & 13.16\\
  & Sapphire & 1.00$'$ & 12.68 & 14.79 & 12.96 \\
  & NUVN2 & 1.67$'$ & 11.07 & 13.41 & 11.66 \\
  \hline
   \multirow{3}{*}{NGC4147} & BaF2 & 0.40$'$& 15.81 & 17.65 & 15.70 \\
   & Sapphire & 0.41$'$ & 15.72 & 17.58 & 15.60 \\
   & Silica & 0.40$'$ & 15.40 & 17.29 & 15.26 \\
   \hline
   \multirow{3}{*}{NGC5053} & BaF2 &- & - & 15.37 & 14.96\\
   & NUVB15 & - & - & 13.74 & 13.34 \\
   & NUVB4 & - & - & 13.39 & 13.04 \\
   \hline
  \multirow{4}{*}{NGC2808} & BaF2 & 1.39$'$& 11.73 & 14.09 & 12.25 \\ 
  & Sapphire & 1.62$'$ & 11.58 & 14.02 & 12.20 \\
  & Silica15 & 2.55$'$ & 10.94 & 14.33 & 12.27 \\
  & NUVN2 & 2.35$'$ & 9.95 & 14.42 & 12.18 \\
  \hline
 \multirow{5}{*}{NGC1261} & Sapphire & - & - & 16.25 & 15.44 \\  
 & Silica & - & - & 15.67 & 14.80 \\
 & NUVB15 & 0.90$'$ & 13.09 & 14.21 & 13.34 \\
 & NUVB13 & 1.67$'$ & 12.46 & 13.99 & 13.11 \\
 & NUVB4 & 1.58$'$ & 11.99 & 13.52 & 12.63 \\
 \hline
  \multirow{4}{*}{NGC5466} & CaF2 & - & - & 15.87 & 15.09 \\
  & Sapphire & - & - & 15.51 & 14.71 \\
  & NUVB13 & 3.93$'$ & 12.64 & 13.90 & 13.23 \\
  & NUVB4 & 3.08$'$ & 11.65 & 13.45 & 12.77 \\
  \hline
  \multirow{3}{*}{NGC288} & CaF2 & 0.91$'$& 14.41 & 13.95 & 13.32 \\ 
  & Sapphire & - & - & 13.74 & 13.10 \\
  & NUVN2 & 2.92$'$ & 11.33 & 12.36 & 11.66 \\
  \hline
 \multirow{2}{*}{NGC362} & CaF2 & 3.33$'$& 10.87 & 13.89 & 11.76 \\ 
 &  NUVB4 & 4.33$'$ & 10.24 & 13.52 & 11.28 \\
 \hline
 \multirow{3}{*}{NGC1851} & CaF2 & 0.82$'$& 13.34 & 16.12 & 13.56 \\ 
 & Sapphire & 0.79$'$ & 12.99 & 15.79 & 13.19 \\
 & NUVN2 & 3.00$'$ & 10.51 & 13.57 & 11.45 \\
  
\hline
\end{tabular}
\end{table}

\subsection{Open Clusters} \label{sec:oc}

UVIT has observed eight open clusters during 2016$-$17. Since open clusters are not as dense as globular clusters, we could properly resolve all sources present in the open cluster fields. The UV sources of open clusters are flagged with flag {\it ID} = 3 in the catalog.

We cross-matched the UV sources of the open clusters with the {\it Gaia} catalog of open clusters \citep{oc_cantat} using a match-radius of 1.5$''$ and selected the cluster member UV sources having a membership probability greater than 70\%. We correct the UV magnitudes for extinction using the extinction curves provided in \cite{cardelliext}. The corrected magnitudes are converted to absolute magnitudes using the cluster distance modulus in \cite{oc_cantat}.

The UV-optical CMDs using Gaia G and UVIT filters are shown in the bottom panel of \autoref{fig:gc-gaia-hb}. We use the BaSTI isochrones shown as the black solid, dotted, and dashed lines to trace the loci of the stars in \autoref{fig:gc-gaia-hb}. We vary age and metallicity parameters as the UV-optical colors are affected by both parameters. The age of the isochrones ranges from 29 Myr to 7 Gyr, while the metallicities ([Fe/H]) vary between $-$0.42 and 0.42. The age and metallicity parameters are taken from \cite{oc_cantat} for all the clusters except for Be67 and NGC663, which were used from \citet{oc_dias}. We separate different evolutionary phases of the stars based on the Gaia-optical CMD plots. The column `Stype' in the open cluster catalog indicates the evolutionary phases of the stars. We find that main-sequence stars (`Stype'=ms) are the most dominant stellar population in open clusters in both FUV and NUV filters, while the red-giant branch stars (`Stype'=rgb) are mostly seen in NUV filters, with an exception for the M67 cluster, where we find 2 RGB stars in FUV filters. The number of BS (`Stype'=bs) and RGB stars detected in different clusters and their properties (age, metallicities, distance modulus) are given in \autoref{tab:oc-details}. We found a few peculiar stars, which we verified using the SIMBAD database, in which two are spectroscopic binaries (`Stype'=sb), one each in the cluster NGC188 and King2 (Sapphire $-$ G $\sim$0 mag), five hot sub-dwarf stars in NGC6791 (NUVB4 $-$ G $\sim$0 mag), and one blue supergiant star in NGC663 with FUV $-$ G color $\sim$1.2 mag.

\begin{table}
		\centering
		\caption{List of open clusters observed during 2016 $–$ 2017. BS and RGB are the number of blue straggler and red giant branch stars detected in the open clusters. The values of age, metallicity, and distance modulus (d$_{mod}$) are taken from \cite{oc_cantat, oc_dias}.}
	   	\label{tab:oc-details}
	\begin{tabular}{cccccccc} % four columns, alignment for each
\hline
			\hline
 \multirow{2}{*}{Name}  & Age & [Fe/H] & \multirow{2}{*}{d$_{mod}$} & \multirow{2}{*}{Filter} & \multirow{2}{*}{BS} & \multirow{2}{*}{RGB} \\
 &    (Gyr) & (dex) &   \\

\hline

 \multirow{4}{*}{King2}  & \multirow{4}{*}{4.07} & \multirow{4}{*}{-0.42} & \multirow{4}{*}{14.15} & CaF2 & 5 & -\\
 & & & & Sapphire &  5 & -\\
 & & & & NUVB15 & 2 & - \\
 & & & & NUVN2 & 10 & 1 \\
 \hline
\multirow{2}{*}{Be67}  & \multirow{2}{*}{1.26} & \multirow{2}{*}{+0.02} & \multirow{2}{*}{11.73} & Silica15  & - & - \\
& & & & NUVB13 & - & - \\
\hline
\multirow{3}{*}{NGC6791} & \multirow{3}{*}{6.31} & \multirow{3}{*}{+0.42} & \multirow{3}{*}{13.13} & Silica & 1 & -\\
 & & & & NUVB13 &  11 & - \\
 & & & & NUVB4 & 22 & - \\
 
 \hline
 \multirow{4}{*}{NGC7789}& \multirow{4}{*}{1.55} & \multirow{4}{*}{+0.02} & \multirow{4}{*}{11.61}  & Sapphire & 11 & - \\
 & & & & NUVB13 &  10 & 21 \\
 & & & & NUVB4 & 10 &  105 \\
  & & & & NUVN2 &  10 & 74\\
 \hline
 \multirow{3}{*}{NGC663} & \multirow{3}{*}{0.03} & \multirow{3}{*}{-0.125} & \multirow{3}{*}{12.35} & Sapphire & - & -\\
  & & & & NUVB4 & - &  - \\
  & & & & NUVN2 &  - & -\\
  \hline
 \multirow{5}{*}{NGC188} & \multirow{5}{*}{7.08} & \multirow{5}{*}{+0.12} & \multirow{5}{*}{11.15} & CaF2 & 8 & -  \\
 & & & & Sapphire & 5 &  - \\
 & & & & Silica & 5 &  - \\
 & & & & NUVB15 & 15 &  - \\
 & & & & NUVN2 & 19 &  39 \\
 \hline
\multirow{3}{*}{M67}  & \multirow{3}{*}{4.27} & \multirow{3}{*}{+0.03} & \multirow{3}{*}{9.75} & CaF2 & 8 & 2  \\
& & & & BaF2 & 6 &  2 \\
& & & & Sapphire & 8 &  2 \\
\hline
NGC2477 & 1.12 & +0.07 & 10.8 & NUVB4 & 5 & 72 \\
\hline

\end{tabular}
\end{table}

\subsection{Gaseous Nebulae} \label{sec:gneb}

\begin{figure*}
    \centering
 
     \includegraphics[width=0.32\textwidth]{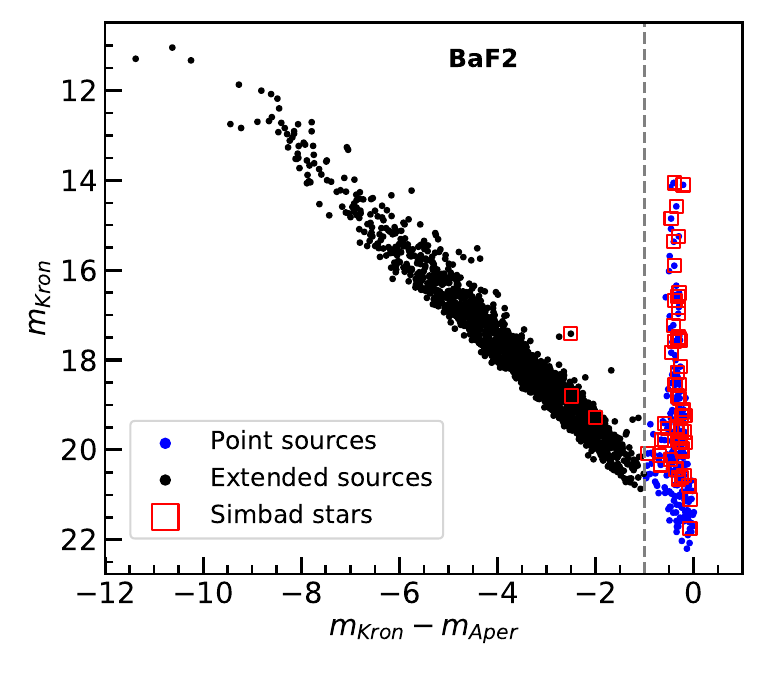}
    \includegraphics[width=0.32\textwidth]{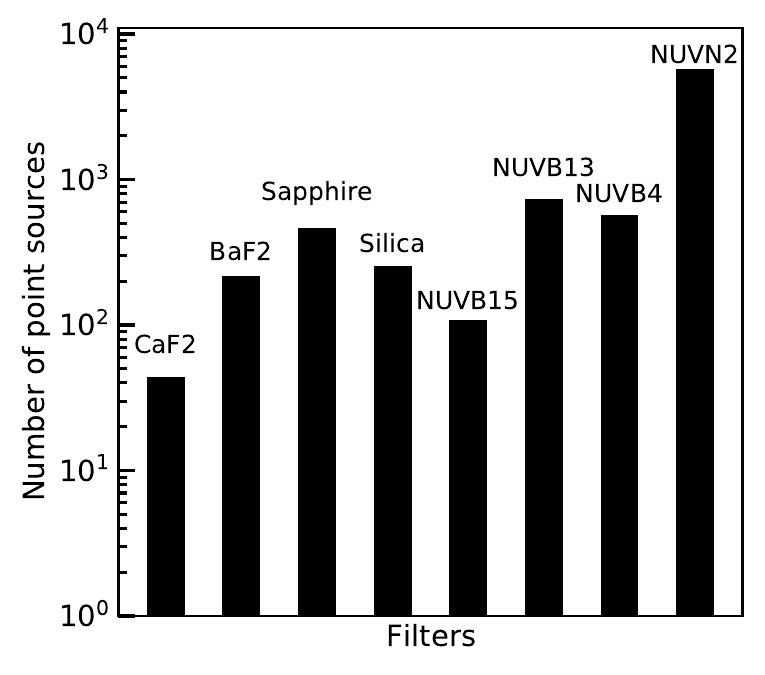}
    \includegraphics[width=0.32\textwidth]{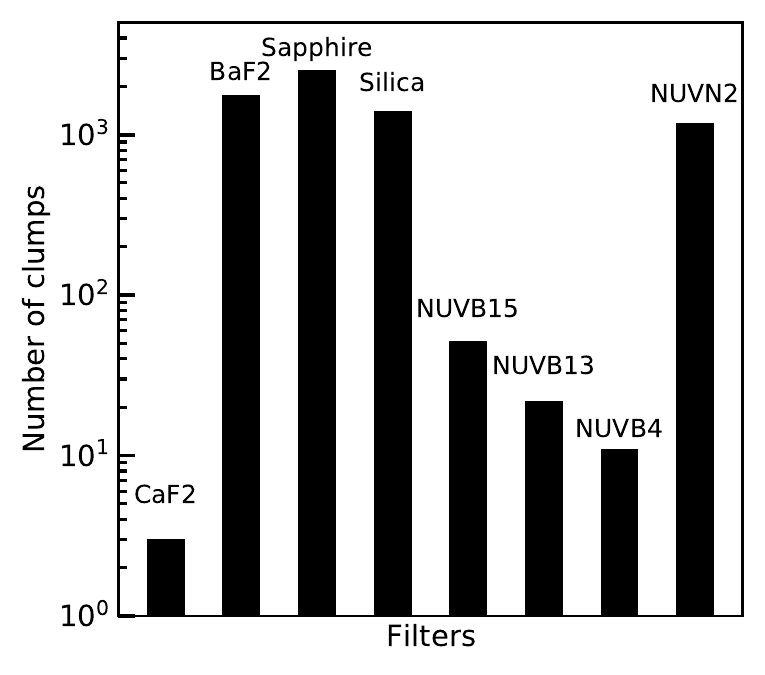}
    \caption{Left: Magnitude difference between Kron elliptical magnitudes and the aperture magnitude (aperture radius = 3 pixels) for the sources observed in the BaF2 filter. The red boxes represent the confirmed stars from the SIMBAD catalog. We find that 90\% of the stars have the $\rm m_{Kron} - m_{Aperture}\ge-$1 mag. The sources with the same magnitude difference are classified as point sources (blue solid circles) in gaseous nebulae and the rest as extended clumps (black solid circles). Middle: The histogram plot of the number of point sources in different filters of gaseous nebulae fields. Right: The histogram plot of the number of clumps in different filters of gaseous nebulae fields.}
    \label{fig:gas-neb}
\end{figure*}

UVIT has observed 11 gaseous nebulae during the 2016$-$2017. We have shown the FUV and NUV images of one of the gaseous nebulae, NGC 6995, observed by UVIT in the middle panel of \autoref{fig:sample-field}. The NUV image shows both point sources as well as clump-like gaseous emissions, while the FUV image mostly includes clumps. We identify UV-emitting regions (clumps) and point sources from 11 gaseous nebulae under UVIT observation. All the sources detected in the gaseous nebulae field are flagged with flag {\it ID} 4.

The gaseous nebulae fields have two types of detection by SE: the UV-bright regions and the field stars. We use the difference between kron magnitude and fixed aperture magnitude (r = 3 pixels (1.2$''$)) to segregate the detected extended and point-like sources. We consider sources with differences in kron and aperture magnitudes less than $-$1.0 mag as point sources and the rest of the sources as extended sources ($\rm m_{Kron} - m_{Aperture}>$ -1). The `Stype' column in the gaseous nebulae catalog indicates the point (`Stype'=1) and extended sources (`Stype'=0). We found 60 point sources for the observed gaseous nebulae from the SIMBAD database, which are over-plotted with red squares in the left panel of \autoref{fig:gas-neb}. It is clear from the figure that all these sources have ($\rm m_{Kron} - m_{Aperture}$) $>$ -1.0 mag. We have also displayed all the point sources (218) of gaseous nebulae detected in the BaF2 filter of UVIT in blue color. The middle and right panels of \autoref{fig:gas-neb} show the number of point sources and the clumps observed in different filters of UVIT. We report more than 2500 (450) and 1100 (5500) UV-emitting regions (point sources) in the Sapphire and NUVN2 filters, respectively.

\subsection{Planetary Nebulae} \label{sec:pne}

UVIT has observed 11 planetary nebulae in 13 FoVs during 2016$-$17. 
Previously, \citet[GUVPNcat,][]{Galexpncat} used Hong Kong/AAO/Strasbourg H$\alpha$ (HASH) and GALEX observations to identify 671 unique planetary nebulae in the Galaxy. Their result consists of magnitude measured in seven different apertures and information on their shapes (major [R$_{maj}$] and minor [R$_{min}$] axes of the nebulae). 

We compute integrated FUV and NUV magnitudes for 11 planetary nebulae (PNe) in our sample using elliptical apertures with dimensions equivalent to R$_{maj}$ and R$_{min}$ given in GUVPNcat. \autoref{tab:pne_details} lists the details about the planetary nebulae observed in UVIT DR1, their coordinates, size information, morphological type, and integrated magnitudes in different filters. Our sample consists of two round (R), five elliptical (E), and four bipolar (B) nebulae. The sources within the elliptical apertures were assigned Flag {\it ID} = 5, whereas the remaining sources outside were given Flag {\it ID} = 0. We highlight \citet{2020pnreport} have previously analyzed the UVIT observations of three planetary nebulae present in UVIT DR1.

\begin{table*}
		\centering
		\caption{Integrated magnitudes of the planetary nebulae (PNe) along with their semi-major (R$_{maj}$) and semi-minor (R$_{min}$) axes and morphological classifications. Magnitudes are given for whichever filters the observation exists.}
	   	\label{tab:pne_details}
\begin{tabular}{lcccccccccccl}
\hline
    Name & RA & DEC & R$_{maj}$ & R$_{min}$ & class & BaF2 & Sapphire & Silica & NUVB15 & NUVB13 & NUVB4 & NUVN2  \\
    %&&&($'$)&($'$)&&\\
    \hline
Abell21 & 112.2613 & 13.2469 & 375.0 & 257.5 & E & - & 11.79 & 11.83 & - & 11.73 & 11.43 & -\\
  Abell30 & 131.7230 & 17.8793 & 63.5 & 63.5 & R & 13.32 & 13.13 & 12.93 & 13.33 & - & - & 13.65 \\
  H4-1 & 194.8658 & 27.6363 & 1.35 & 1.35 & E & 17.77 & - & - & - & 17.65 & 18.18 & -\\
  LoTr5 & 193.8906 & 25.8918 & 262.5 & 255.0 & E & - & 12.70 & 12.22 & - & 13.17 & 12.71 & -\\
  NGC3587 & 168.6990 & 55.0190 & 104.0 & 101.0 & R & - & 12.26 & 12.22 & - & - & - & -\\
  NGC40 & 3.2543 & 72.5220 & 28.0 & 17.0 & B & - & - & - & 13.18 &- & - & 12.47\\
  NGC6302 & 258.4354 & -37.1032 & 195.0 & 195.0 & B & 12.47 & 12.42 & 12.48 & 12.77 & - & - & 11.80\\
  NGC7293 & 337.4106 & -20.8371 & 485.0 & 367.5 & B & - & 9.97 &9.94 &- & 9.96& 9.94 & -\\
  NGC1514 & 62.3208 & 30.7760 & 94.0 & 91.0 & E & - & - & -& - & 13.95 & 13.47 & -\\
  NGC2440 & 115.4808 & -18.2087 & 36.0 & 36.0 & B & - & 12.80 & 12.76 & - & 13.55 & 13.44 &-\\
  NGC7094 & 324.2207 & 12.7887 & 51.25 & 49.5 & E & - & 13.23 & 12.73 & - & 14.02 & 13.89 & -\\
\hline
\end{tabular}
\end{table*}

\subsection{Magellanic Cloud} \label{sec:mc}

There are 13 UVIT observations of different regions of the Magellanic clouds during 2016$-$17. These observations comprise seven fields of Small Magellanic Cloud (SMC), four fields of Large Magellanic Cloud (LMC), and two fields observing the Magellanic bridge, which lies in between the SMC and LMC with PIDs A04\_066 and T01\_145. We followed the source extraction procedure mentioned in \autoref{sec:source_detection} and extracted most point sources from the observations.  All the sources in the Magellanic fields are flagged with Flag {\it ID} = 6. We have also identified five clusters in Magellanic Clouds detected in various filters of UVIT, which were confirmed after cross-matching our catalog with the SMC cluster catalog of \citet{smc_clusters}. However, these clusters are not provided in our UVIT catalog.

\subsection{Nearby Galaxies}
\label{sec:NG}

We use the HyperLEDA catalog of galaxies \citep{hyperleda} to identify the nearby galaxies in the UVIT observations. We impose an essential criteria on the semi-major axis of the galaxies, R$_{maj}$ $>$  1$'$, to segregate nearby galaxies from HyperLEDA catalog \citep{2007galexnbg}. We cross-match the HyperLEDA nearby galaxies with the central coordinates of the FoVs observed in UVIT DR1 to identify the set of UVIT observed nearby galaxies. A cross-matching radius of 14$'$ (equivalent to the FoV radius of UVIT) is used.  We find that a total of 181 nearby galaxies are observed in different filters of UVIT during 2016$-$2017. The right panel of \autoref{fig:sample-field} shows a nearby spiral galaxy along with other background and foreground sources in FUV and NUV. We highlight the prominent spiral arms, which consist of several star-forming clumps detected by UVIT. 

We detect 85 galaxies (46 FoVs) in BaF2, 54 galaxies (34 FoVs) in CaF2, 42 galaxies (26 FoVs) in Silica, and 39 galaxies (25 FoVs) in Sapphire filters. In NUV, we report 80 galaxies (42 FoVs) in NUVB4, 73 galaxies (40 FoVs) in NUVB13, 62 galaxies (34 FoVs) in Silica15, 58 galaxies (30 FoVs) in NUVB15, and 49 galaxies (26 FoVs) in NUVN2 filters. We detect local galaxies with sizes ranging from a few arcmin to more than half of the FoV. Out of these galaxies, 55 are spiral, 13 are elliptical, and 2 are irregular galaxies in the BaF2 filter. In the Silica15 band, 35 galaxies are spiral, 11 are elliptical, and 2 are irregular galaxies. The distance of these galaxies ranges from 0.004 Mpc to 266 Mpc, and the value of their R$_{maj}$ varies between 1$'$ and 29.85$'$. 

\begin{figure}
    \centering
    \includegraphics[width=0.49\textwidth]{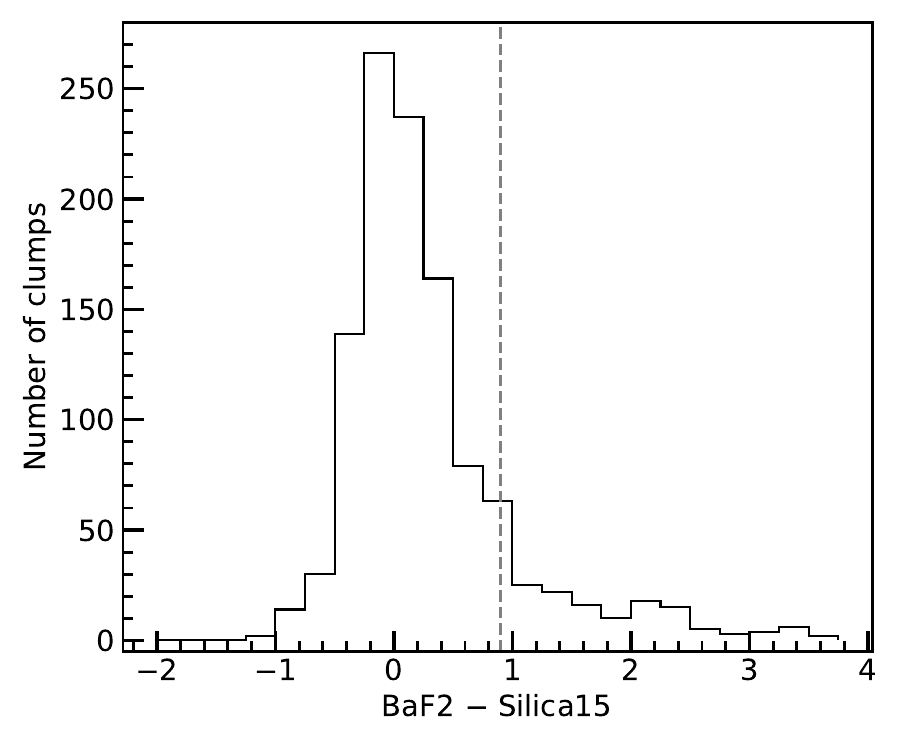}
    \caption{FUV$-$NUV histogram for clumps commonly detected in BaF2 and Silica15 filters. The gray dashed line at color FUV$-$NUV $=$ 0.9 mag segregates young and old stellar population-dominated galaxies \citep{2007galexnbg}.}
    \label{fig:nbg-color-hist}
\end{figure}

Due to their proximity and the good spatial resolution of UVIT, we were able to resolve the star-forming clumps inside the nearby galaxies.  The clumps present within a defined elliptical region around the center of the galaxies are considered part of it. The parameters of these ellipses, i.e., R$_{maj}$, R$_{min}$ (length of semi-minor axis of the galaxy), and the position angle are taken from the HyperLEDA catalog. The estimated magnitudes of the clumps resolved within the defined elliptical apertures, mimicking the optical dimensions taken from HyperLEDA, are further analyzed in this work. These clumps are identified in the catalog with Flag {\it ID} = 7. The sources lying outside the elliptical aperture of a galaxy are given Flag {\it ID} = 0. Combining altogether, 18,034 clumps are observed in all the FUV filters, and 41,033 clumps are obtained in NUV filters. We show the Galactic dust extinction corrected FUV$-$NUV color histogram for all the clumps commonly detected in BaF2 and Silica15 filters in \autoref{fig:nbg-color-hist}. The FUV$-$NUV color peaks at 0.9 mag for a large sample of galaxies, and the galaxies transition from young to old-type galaxies \citep{2007galexnbg}. We detected 972 blue and 155 red clumps in both BaF2 and Silica15 filters. This implies that the clumps in our sample incline towards younger and star-forming. A detailed analysis of the properties of star-forming clumps will be carried out in an upcoming work.

\section{Summary and Conclusions} \label{sec:summary}

We summarize the contents of the catalog as follows:

\begin{itemize}
    \item We provide the catalog, UVIT DR1, consisting of UVIT sources observed in the first two years of observation until December 31, 2017. We have 291 unique fields after removing the fields with multiple entries and fields with no bright sources. Of these, 221 fields have FUV and NUV observations, 36 have only FUV observations, and 34 have only NUV observations. The exposure times of the observations range from 120 s to 68 ks. 
    \item We use {\sc ccdlab} pipeline for data reduction and the {\it Gaia} DR2 catalog for astrometry. The uncertainties in the astrometry are between 0.1$''$ and 0.5$''$.
    \item We estimated the background mean and RMS error by masking the sources above 2$\sigma$ and placing at least 1000 boxes of size 21 $\times$ 21 pixels. We calculated the 3$\sigma$ and 5$\sigma$ detection limits using \autoref{eq:detect_limit}. The 5$\sigma$ detection limit of a typical 100 s exposure is 26.7 (25.6) mag in the Silica 15 (BaF2) filter.
    \item We use PSFEx to estimate the average FWHM of the image. This value ranges from $1.0''-2.0''$ for FUV and $0.8''-1.5''$ for NUV filters.  
    \item We use a source extractor for source detection.  In the catalog, we provide a variety of fixed and flexible aperture fluxes and magnitudes, i.e., aperture magnitudes for three circular apertures of radii 3, 12, and 15 pixels (1.25$''$, 5.00$''$, and 6.26$''$), Kron-like elliptical aperture magnitude, PSF magnitude, etc. We also provide the physical properties of the sources, such as their elliptical sizes, orientation, ellipticity, and FWHM. The columns included in the catalog are given in \autoref{sec:appendix}, and the catalog will be available online. 
    
    \item We verify our deblending parameters using several globular clusters observed by UVIT. We discuss the measures taken to remove various artifacts caused by diffraction spikes, edge artifacts, satellite trails, bright galaxies, etc.

     \item We cross-match field sources (Flag {\it ID} = 0) in UVIT DR1 with the {\it GALEX} AIS catalog. We found that more than 90\% of matched sources are single-matched, while the rest are multiple-matched. Some of the GALEX sources do not have UVIT counterparts. These sources lie at the edge of UVIT FoV (normalized exposure $\le$90\%). The enhanced resolution added with deep exposures taken by UVIT pushes the detection of sources to fainter limits compared to {\em GALEX}. As a result, our catalog presents the first detection of several sources at UV wavelengths. We determine the magnitude difference between UVIT and {\em GALEX} magnitudes to be $\le$1 mag for FUV filter while greater than 1 for NUV filters towards the brighter end. At fainter end, the difference is greater than 1 mag due to deeper observations of UVIT.

    \item We also match UVIT DR1 with the {\it Gaia} DR3 catalog in a comparative study with optical observations. We could find optical counterparts for more than 90\% of the sources in the {\it Gaia} DR3 catalog. Most of these sources are single-matched, but a few multiple-matched sources are also found. We find the percentage of spurious matches for sources observed in non-crowded field to be $\sim$0.1$-$0.15\% of the total matched sources in NUV filters but we did not find any spurious matches for FUV filters. For the combined UVIT DR1 catalog, we found spurious matches for $\sim$1$-$2\% of the total matched sources in NUV filters and $\sim$1$-$1.5\% of the total matched sources in FUV filters.

    \item We cross-match M31 sources in UVIT DR1 with \cite{Leahy2020} to find that only point sources could be matched, while there is a discrepancy in the identification of clumps within the spiral arms of M31. The difference in source detection parameters and magnitude calculation between the catalogs could be the reason for the discrepancy. The magnitude difference in non-crowded region for a single FoV is $\sim$0.5 mag while in crowded region it increases to $\sim$2 mag.
    
    \item The catalog covers an area of $\sim$58.2 deg$^2$. The number of sources obtained in each filter is given in \autoref{tab:source_numbers}.

      \item The non-crowded field sources include all the sources, excluding the large angular-sized objects such as nearby galaxies, star clusters, and nebulae systems.  We have used magnitude histogram plots to determine
the depth at which most sources are detected.
      %We approximately determine the depth in the detection of the sources in the catalog using the magnitude histogram plots. 
      In FUV filters, the depth in detection limit for CaF2, BaF2, Sapphire, and Silica is at 22.75 mag, 22.25 mag, 21.75 mag, and 20.75 mag, whereas in NUV filters, depth in detection limit for Silica15, NUVB15, NUVB13, NUVB4, and NUVN2 filters is at 23.75 mag, 20.75 mag, 22.25 mag, 21.25 mag, and 20.25 mag.
      
    \item All the unique sources in the catalog have the $N_{epoch}$ flag as `0\_0' while the multi-epoch observations have non-zero $N_{epoch}$ values. The number of multi-epoch sources obtained in each filter is given in \autoref{tab:source_numbers}.

    \item The SE does not resolve the central region of the globular clusters. In the catalog, we give a magnitude (m$_{R}$) within the radius (R) outside of which all the sources are resolved. The sources inside this radius are removed. In \autoref{tab:gc-core-mag}, we give the magnitude within the cluster's half-light radius and core radius. We find that the most dominant stars in GCs are the BS, HB, and post-HB stars. We detect 500 HB stars in CaF2, 211 in BaF2, 758 in Sapphire, and 134 in Silica filters. We find 45 BS stars in CaF2, 7 in BaF2, 37 in Sapphire, and 3 in Silica. We identify 6 post-HB stars in CaF2, 14 in BaF2, 19 in Sapphire, and 5 in Silica.
    
    \item We cross-match the open cluster sources in UVIT DR1 with \cite{oc_cantat} and select the cluster members of the stars. Based on the position in the UV-optical CMD plot, we have identified different evolutionary phases of open clusters such as main-sequence, main-sequence turn-off, red-giant branch, and sub-giant branch stars, as shown in \autoref{fig:gc-gaia-hb}. We found a few peculiar sources and verified the sources by cross-matching with SIMBAD. These sources include two spectroscopic binary stars (one each in NGC188 and King2 clusters), five hot-subdwarf stars in NGC6791, and one blue-supergiant star in NGC663.
    
    \item The gaseous nebulae fields have clumps and point sources. We use the difference between kron-magnitude and fixed aperture magnitude (radius $=$ 3 pixel (1.2$''$)) to separate the point sources from the extended clump-like sources. We find that more than 90\% of the stars (confirmed from SIMBAD) have m$_{Kron}$ -m$_{Aperture} \ge -1$ mag (shown in the left panel of \autoref{fig:gas-neb}). In the catalog, we have more than 2500 (450) and 400 (5500) UV-emitting regions (point sources) in Sapphire and NUVN2 filters.
    
    \item  We cross-match UVIT DR1 with GUVPNcat \citep{Galexpncat} to identify the planetary nebulae in the catalog. We use the major (R$_{maj}$) and minor (R$_{min}$) axes of the nebulae given in GUVPNcat to compute the UV integrated magnitude of the planetary nebulae in different filters of UVIT. Our sample consists of two round (R), five elliptical (E), and four bipolar (B) nebulae. We provide the integrated magnitude of the nebulae and their morphological type in \autoref{tab:pne_details}.

    \item For Magellanic cloud observations, we extract the point sources observed in LMC, SMC, and Magellanic Bridge. We identified and confirmed five clusters by cross-matching with the SMC cluster catalog of \citet{smc_clusters}.

    \item We use the Hyperleda catalog of galaxies to separate the nearby galaxies having a semi-major axis greater than 1$^{\prime}$. We found 78 galaxies (37 fields) in BaF2, 48 galaxies (35 fields) in CaF2, 38 galaxies (23 fields) in Silica, and 37 galaxies (22 fields) in Sapphire filters. In NUV, we found 76 galaxies (42 fields) in NUVB4, 69 galaxies (37 fields) in NUVB13, 59 galaxies (34 fields) in Silica15, 51 galaxies (26 fields) in NUVB15, and 46 galaxies (22 fields) in NUVN2 filters. Most of these galaxies are spiral, and the distances of the galaxies range from $0.004 - 266$ Mpc. 
    
\end{itemize}

%\begin{acknowledgement}
This publication uses the data from the {\it AstroSat} mission of the ISRO, archived at the Indian Space Science Data Center (ISSDC). ACP and SP acknowledge the support of Indian Space Research Organisation (ISRO) under {\it AstroSat} archival Data utilization program (No. DS\_2B-13013(2)/1/2022-Sec.2). ACP also thanks Inter University centre for Astronomy and Astrophysics (IUCAA), Pune, India for providing facilities to carry out his work. We thank the referee
for careful reading and useful comments and suggestions, which helped us to improve the manuscript.
%\end{acknowledgement}

\bibliography{references}{}
\bibliographystyle{aasjournal}

\appendix %\label{sec:appendix}
	
\section{Additional Information} \label{sec:appendix}

\begin{center}

\begin{longtable}{|l|l|l|}
\caption{Details of UVIT catalog columns.} \label{tab:catalog columns}

 \\

\hline \multicolumn{1}{|c|}{\textbf{Keyword}} & \multicolumn{1}{c|}{\textbf{Description}} & \multicolumn{1}{c|}{\textbf{Unit}} \\
\hline 
\endfirsthead

\multicolumn{3}{c}%
{{\bfseries \tablename\ \thetable{} -- continued from previous page}} \\
\hline \multicolumn{1}{|c|}{\textbf{Keyword}} & \multicolumn{1}{c|}{\textbf{Description}} & \multicolumn{1}{c|}{\textbf{Unit}} \\ \hline 
\endhead

\hline %\multicolumn{3}{|r|}{{Continued on next page}} \\ \hline
\endfoot

\hline %\hline
\endlastfoot

PID & Proposal ID of the image & - \\
TID & Target ID of the image & - \\
PIDTID & Field identification number of format PID\_TID & - \\
UID & Source identification number of format PID\_TID\_NUMBER & - \\
RA\_J2000 & Right Ascension of the source & degree\\
DEC\_J2000 & Declination of the source & degree\\
X\_IMAGE & Physical x-coordinate of the source & pixel\\
Y\_IMAGE & Physical y-coordinate of the source & pixel\\
  NUMBER & Source number assigned by SE & -\\

  MAG\_ISO & Isophotal magnitude & mag \\
  MAGERR\_ISO & Error in isophotal magnitude & mag \\

  MAG\_AUTO & Kron magnitude of the source & mag \\
  MAGERR\_AUTO & Error in kron magnitude & mag \\ 

  SNR\_WIN & Signal-to-noise ratio of the source &-\\
  KRON\_RADIUS & Kron aperture radius&- \\

  BACKGROUND & Local background of the source & count\\

 A\_IMAGE & Semi-major axis of the elliptical aperture & pixel\\
 B\_IMAGE & Semi-minor axis of the elliptical aperture & pixel\\
 THETA\_IMAGE & Position angle of the source & degree\\
  A\_WORLD & Semi-major axis of the elliptical aperture & degree\\
 B\_WORLD & Semi-minor axis of the elliptical aperture & degree\\
 THETA\_WORLD & Position angle of the source & degree\\
FLUX\_RADIUS & Radius enclosing half of the total flux of the object & pixel\\ 
 FWHM\_IMAGE & FWHM of the source & pixel\\

ELONGATION & A\_IMAGE/B\_IMAGE & -\\
ELLIPTICITY & 1 - B\_IMAGE/A\_IMAGE & -\\
CLASS\_STAR & Star/Galaxy classifier & - \\

  MU\_MAX & Peak surface brightness above background & mag \\

  MAG\_APER\_3 & Aperture magnitude within   aperture radius of 3 pixels & mag \\
  MAGERR\_APER\_3 & Error in aperture magnitude   within aperture radius of 3 pixels & mag  \\
 
    MAG\_APER\_12 & Aperture magnitude within aperture radius of 12 pixels & mag \\
  MAGERR\_APER\_12 & Error in aperture magnitude within aperture radius of 12 pixels & mag \\

    MAG\_APER\_15 & Aperture magnitude  within aperture radius of 15 pixels & mag  \\
  MAGERR\_APER\_15 & Error in aperture magnitude within aperture radius of 15 pixels  & mag \\

  IRAF\_MAG\_PSF & Magnitude obtained after fitting Gaussian function & mag\\
  IRAF\_MAG\_PSF\_COR & PSF magnitude after applying aperture and saturation corrections & mag\\
  IRAF\_MAGERR & Error in PSF magnitude & mag\\
  CHI & Goodness of PSF-fitting & -\\
  SHARPNESS & Sharpness of PSF-fitting & -\\

  Exp\_Map & Exposure value of the source & - \\
  Exp\_Time & Exposure time of the image & seconds \\
  PSF\_FWHM & PSF FWHM of the image & pixel\\
  RA\_fc & Right Ascension coordinate of center of the FoV & degree\\
  DEC\_fc & Declination coordinate of center of the FoV & degree \\
  cdist & Distance of the source from the center of the FoV& arcmin \\

\end{longtable}

\begin{longtable}{|l|l|l|}
\caption{Details of various Flags mentioned in the catalog} \label{tab:catalog columns}

 \\

\hline \multicolumn{1}{|c|}{\textbf{Keyword}} & \multicolumn{1}{c|}{\textbf{Description}} & \multicolumn{1}{c|}{\textbf{Unit}} \\
\hline 
\endfirsthead

\multicolumn{3}{c}%
{{\bfseries \tablename\ \thetable{} -- continued from previous page}} \\
\hline \multicolumn{1}{|c|}{\textbf{Keyword}} & \multicolumn{1}{c|}{\textbf{Description}} & \multicolumn{1}{c|}{\textbf{Unit}} \\ \hline 
\endhead

\hline %\multicolumn{3}{|r|}{{Continued on next page}} \\ \hline
\endfoot

\hline %\hline
\endlastfoot

Flagid & Flag to identify the field type (Flag {\em ID} for M31 galaxy fields: 1; globular clusters: 2; open clusters: 3; & -\\ & gaseous nebulae: 4; planetary nebulae: 5; Magellanic clouds: 6; Nearby galaxies: 7)  & -\\
  Nepoch & Flag to identify the single and multi-epoch sources& -\\
  Stype (GC) & Evolutionary phases identified in globular clusters (BHB stars: bhb; post-HB: phb; BS stars: bss).  & -\\
  Stype (OC) & Evolutionary phases identified in  open-clusters (BS stars: bss; RGB: rgb, main-sequence stars: MS;& -\\ & main-sequence turn-off: MSTO; spectroscopic binaries: SB; sub-dwarf stars: sub-dwarf; & - \\
  & sub-giant branch stars: SGB; blue-supergiant stars: BSG). & -\\
  Stype (GNeb) & Source type in gaseous nebulae (point sources; 1; extended sources: 0) & -\\
  NS/G & Probable star or galaxy (point sources; 1; extended sources: 0) & - \\

\end{longtable}

\end{center}

%% This command is needed to show the entire author+affiliation list when
%% the collaboration and author truncation commands are used.  It has to
%% go at the end of the manuscript.
%\allauthors

%% Include this line if you are using the \added, \replaced, \deleted
%% commands to see a summary list of all changes at the end of the article.
%\listofchanges

\end{document}